\documentclass[12]{iopart}
\usepackage{amsmathred, latexsym, amssymb, amscd,amsfonts, amsthm, epsfig,color,times,appendix}

\usepackage{setspace}

\usepackage{graphicx}
\usepackage{multirow}
\usepackage{pictex}
\usepackage[english]{babel}

\newtheorem{theorem}{\bf Theorem}[section]

\begin{document}

\title[Rank-based model selection for multiple ions quantum tomography]
{{\sf \bfseries Rank-based model selection for multiple ions quantum tomography}}
\author{ \sf \bfseries  M\u{a}d\u{a}lin Gu\c{t}\u{a}$^{1}$,   Theodore Kypraios$^{1}$ and Ian Dryden$^{2,1}$ } 
\address{$^{1}$School of Mathematical Sciences, University of Nottingham,
University Park, NG7 2RD Nottingham, United Kingdom}
\address{$^{2}$ Department of Statistics, University of South Carolina, Columbia, SC 29208, USA}
\date{}


\begin{abstract}
The statistical analysis of measurement data has become a key component of many quantum engineering 
experiments. As standard full state tomography becomes unfeasible for large dimensional quantum systems, one needs to exploit prior information and the ``sparsity'' properties of the experimental state in order to reduce the dimensionality of the estimation problem. In this paper we propose model selection as a general principle for finding  the simplest, or most parsimonious explanation of the data, by fitting different models and choosing the estimator with the best trade-off between likelihood fit and model complexity. We apply two well established model selection methods -- the Akaike information criterion (AIC) and the Bayesian information criterion (BIC) -- to models consising of states of fixed rank and datasets such as are currently produced in multiple ions experiments. We test the performance of AIC and BIC on randomly chosen low rank states of 4 ions, and study the dependence of the selected rank with the number of measurement repetitions for one ion states. We then apply the methods to real data from a 4 ions experiment aimed at creating a Smolin state of rank 4. The two methods indicate that the optimal model for describing the data lies between ranks 6 and 9, and the Pearson $\chi^{2}$ test is applied to validate this conclusion. Additionally we find that the mean square error of the maximum likelihood estimator for pure states is close to that of the optimal over all possible measurements.

\end{abstract}
\clearpage

\tableofcontents

\title[Rank-based model selection for multiple ions quantum tomography]
\bigskip

\section{Introduction}

Recent years have witnessed significant progress in the engineering and control of quantum 
systems \cite{NatureInsight,Haroche&Raimond,Dowling&Milburn}. From the preparation of exotic quantum states \cite{Smithey,Resch,Zavatta,Haffner} to the implementation of accurate quantum protocols \cite{Altepeter,OBrien,Riebe,Barreiro} experimentalists are confronted with the problem of reconstructing such mathematical objects \emph{statistically}, from the outcomes of repeated measurements. The theoretical and experimental challenges have stimulated the development of a large array of new methods at the boundary between quantum theory and statistics: state estimation (or tomography) \cite{BlumeKohout,BlumeKohout2,Audenaert&Scheel,Ng,Smolin&Gambetta,Heinosaari}, tomography for incomplete data \cite{Buzek,Teo,Teo&Englert} design of experiments \cite{Smith,Merkel,Nunn}, quantum process and detector tomography \cite{Rahimi,Lundeen} construction of confidence regions (error bars) \cite{BlumeKohout3,Christandl&Renner}, quantum tests \cite{Jupp,Temme} entanglement estimation \cite{LandonCardinal}, 
quantum homodyne tomography \cite{Vogel&Risken,D'Ariano.3,Lvovsky&Raymer}, asymptotic theory 
\cite{Guta&Kahn2,Hayashi&Matsumoto,Audenaert&Szkola}; see also the monographs \cite{Holevo,Helstrom} and the collections of papers \cite{Paris.editor,Hayashi.editor}.

The importance and difficulty of quantum state tomography became evident in the landmark experiment \cite{Haffner} where entangled states of up to 8 ions were created and fully characterised. More recently the same group succeeded in creating entangled states of 14 ions \cite{Monz} but their statistical reconstruction is beyond current computational capabilities! Therefore, there is great interest in alternative methods aimed at reducing the dimensionality of the state estimation problem without making unwarranted or unrealistic assumptions. Among these we mention the development of quantum compressed sensing methods \cite{Gross&Liu&Flammia,Flammia&Gross} which extend the ``classical'' 
$\ell_{1}$-minimisation algorithms \cite{Candes,Tibshirani} to the quantum set-up, and the estimation of many-body states based on lower dimensional families of matrix product states \cite{Cramer1}.  Both methods rely on the Ansatz that the states produced in real experiments are not completely arbitrary, but have some \emph{sparsity} structure that can be exploited for more efficient estimation, e.g. low rank in the first case and finite correlations in the second.

In this paper we propose and investigate a state tomography method  which can also take advantage of the sparsity structure of the state, by adjusting the \emph{rank} of the estimator according to the measurement data. 
However, although it shares with compressed sensing the goal of exploiting sparsity structures, our method is closer to the standard tomography set-up in the sense that it takes as input the dataset consisting of \emph{measurement counts}  rather than \emph{estimates of observables expectations}, and it uses maximum likelihood for determining the estimator of a  given rank. The philosophy of \emph{rank-based model selection} is to choose an estimator which offers a good fit to the data, but in the same time contains a minimal number of parameters (Occam razor principle). For this, we construct a sequence of models consisting of states of fixed rank, and choose the model whose maximum likelihood estimator achieves the best trade-off between fit (likelihood) and model complexity. To quantify the trade-off we use two model selection methods, the Akaike information criterion (AIC) \cite{Akaike} and the Bayesian information criterion (BIC) \cite{Schwarz} which have been used  extensively in model selection problems;  see \cite{Kadane&Lazar,Zucchini} for an introduction to model selection methods, and \cite{Kahn,Usami,Yin} for applications in quantum statistics.

Although the method can be used for an arbitrary measurement set-up, we focus on the statistical model of multiple ions tomography (MIT) \cite{Haffner,Monz}, which constitutes a physically relevant testing ground for tomography of large dimensional systems. We emphasise that model selection does not assume any particular model,  but rather lets the data select the model which gives the most suitable description.This offers the experimentalist an ``honest'' but also minimal estimation framework. The states created in many experiments have a good degree of purity, and therefore one would only need to compute the maximum likelihood over spaces of low rank, rather than full rank matrices. 
Furthermore, the principle of model selection can be applied to other families of models such as matrix product states, which may be more suitable in specific experimental conditions.

The paper is organised as follows. In section \ref{sec.background} we introduce the statistical model of MIT, and discuss some of the existing estimation methods. To gain more insight, in section \ref{sec.pure.states} we investigate the problem of estimating \emph{pure states} and in particular we find that the MIT measurement set-up is quasi-optimal in the sense that the mean norm-two distance squared is only slightly larger than that of the (asymptotically) optimal measurement. 
Section \ref{sec.model.selection} introduces the two rank-based model selection procedures based on the AIC and BIC, and discusses the implementation of the fixed rank models by using the Cholesky decomposition. The methods are applied to three randomly chosen states of ranks 1, 2 and 3 in section \ref{sec.low.rank.states}. We find that both criteria perform very well when the strictly positive eigenvalues are significant relative to the number of measurement samples, and explain this by analysing the asymptotics of the log-likelihood ratio statistic for different models. In section \ref{sec.one.ion} we  investigate the dependence of the selected 
rank on purity and number of measurement repetitions in a one ion study. In section \ref{sec.real.data} we apply BIC and AIC to experimental data provided by Rainer Blatt's group from the University of Innsbruck. We find that the maximum log-likelihood flattens from rank 10 (see Figure \ref{fig.likelihoods}), and the AIC and BIC predict ranks 9 and respectively 6, which capture the 4 significant eigenvalues 
and some of the eigenvalues of order $10^{-2}$ (see Table \ref{table.BIC-AIC-realdata}). 
As additional check, we use the Pearson $\chi^{2}$ statistic to test the hypothesis $H_{0}$ that state has rank at most rank 10, and find that  there is no evidence to reject $H_0$. Section \ref{sec.conclusions} contains a summary of the paper and an outlook for future work.

\section{Background on multiple ions tomography}\label{sec.ion.tomography}
\label{sec.background}

In this section we review the statistical model describing the measurement data collected in multiple ions tomography (MIT) experiments \cite{Haffner,Monz,Barreiro}, and comment briefly on the existing estimation methods, with an emphasis on maximum likelihood estimation.

\vspace{2mm}

The physical system consists of an array of trapped ions whose joint state can be manipulated by means of precisely tuned laser pulses. Since only two electronic energy levels are used for encoding the state, each ion can be describe mathematically as a two level system, so that the joint Hilbert space of $k$ ions is $\mathbb{C}^{2^{k}}$. The state of the system is described by a density matrix $\rho$ on this space, i.e.  a $2^{k}\times 2^{k}$ complex selfadjoint matrix which is positive semidefinite and has trace one. Typically, the goal of the experiment is demonstrate the preparation of a certain target state to a sufficiently high degree of precision.  To validate the result, a large number of preparation-measurement cycles are performed, and the collected measurement data are used to estimate the state produced in the preparation phase. 

In a nutshell, the measurement procedure consists of performing simultaneous Pauli measurements on all ions, each combination of Pauli observables being repeatedly measured $n$ times. More precisely, each measurement is defined by a setting ${\bf  d}$ which specifies which of the 3 Pauli observables $\sigma_{x},\sigma_{y},\sigma_{z}$ is measured for each ion. For instance ${\bf d}:= (x,y,z,z)$ is a 4 ions measurement setting, and in general  for a $k$-ions state there are $3^{k}$ possible settings 
${\bf d}\in \mathcal{D}_{k} :=\{x,y,z\}^{k}$. For each fixed setting, the measurement produces random outcomes 
${\bf s} \in \mathcal{O}_{k}:= \{+1,-1\}^{k}$ with probability distribution 
\begin{equation}\label{eq.proba}
\mathbb{P}_{\rho}({\bf s}|{\bf d}) := {\rm Tr}(\rho P^{\bf d}_{\bf s} )=
\langle e_{\bf s}^{\bf d}|\rho |e_{\bf s}^{\bf d}\rangle ,
\end{equation}
where $P^{\bf d}_{\bf s}$ are one dimensional projections onto the vectors of the orthonormal basis
\begin{equation}\label{eq.basis.d}
|e_{\bf s}^{\bf d} \rangle:= | e_{s_{1}}^{d_{1}} \rangle \otimes\dots \otimes |e_{s_{k}}^{d_{k}}\rangle,
\qquad {\bf s}\in  \mathcal{O}_{k}:= \{+1,-1\}^{k},
\end{equation}
formed by taking tensor products of eigenvectors of the Pauli matrices $\sigma_{d_{1}}, \dots , \sigma_{d_{k}}$:
$$
\sigma_{d}  | e_{s}^{d}\rangle = s\,  | e_{s}^{d}\rangle,  \qquad d\in \{x,y,z\} , \, s\in \{+1, -1\}.
$$
After repeating $n$ times the measurement with setting ${\bf d}$, the data can be summarised by counting the number of times that each possible outcome has occurred.  The probability of a certain set of counts $\{ N({\bf s}|{\bf d}) : {\bf s} \in \mathcal{O}_{k}\}$ is given by the multinomial distribution with probabilites given by \eqref{eq.proba}, so that
\begin{equation}\label{eq.multinomial}
\mathbb{P}_{\rho}( \{ N( {\bf s} |{\bf d} ) : {\bf s} \in \mathcal{O}_{k} \}) =
\frac{n!}{\prod_{s} N( {\bf s} | {\bf d} )!} 
\prod_{\bf s} \mathbb{P}_{\rho} ({\bf s}|{\bf d})^{N({\bf s}|{\bf d}) }, \qquad {\bf d}\in \mathcal{D}_{k}.
\end{equation}
Since any given setting ${\bf d}$ gives information only about the diagonal of the density matrix $\rho$ with 
respect to the basis \eqref{eq.basis.d},  the above procedure is repeated for all possible settings to obtain the complete 
$2^{k} \cdot 3^{k}$ dataset consisting of counts $ \{ N({\bf s}|{\bf d}) :  ({\bf s}, {\bf d}) \in \mathcal{O}_{k} \times \mathcal{D}_{k} \}$ for all outcomes in each setting. As successive preparation-measurement cycles are independent of each other, the distribution over all possible datasets is the product of multinomials
\begin{equation}\label{eq.multinomial.prod}
\mathbb{P}_{\rho}( \{ N( {\bf s} |{\bf d} ) :  ({\bf s}, {\bf d}) \in \mathcal{O}_{k} \times \mathcal{D}_{k}  \}) = 
\prod_{\bf d} \mathbb{P}_{\rho}( \{ N( {\bf s} |{\bf d} ) : {\bf s} \in \mathcal{O}_{k} \}). 
\end{equation}
Let us ponder for a moment on the structure of this statistical model. If no assumption is made on the state, the parameter space is the $(4^{k}-1)$-dimensional convex set  of density matrices $\mathcal{S}_{k}\subset M(\mathbb{C}^{2^{k}})$. We will verify that the above measurement scheme is \emph{informationally complete}, or equivalently that the  parameter $\rho$ is \emph{identifiable} in the sense that there is a one-to-one correspondence between $\rho$ and the probability distribution $\mathbb{P}_{\rho}$ given in \eqref{eq.multinomial.prod}.  
Since $\{\sigma_{x},\sigma_{y},\sigma_{z},\sigma_{0}:=\mathbf{1}\}$ form a basis in the space of $2\times 2$ selfadjoint  matrices, the tensor products
$$
\tilde{\sigma}_{\bf i}:= \frac{1}{2^{k/2}}\sigma_{i_{1}} \otimes \dots \otimes \sigma_{i_{k}} ,\qquad  {\bf i}:= (i_{1},\dots, i_{k})\in \{x,y,z,0\}^{k}
$$
form an orthonormal basis of the space of $2^{k}\times 2^{k}$ selfadjoint matrices with respect to the inner product $\langle A,B \rangle := {\rm Tr}(AB)$. Therefore, any state can be expanded as 
\begin{equation}\label{eq.pauli.fourier}
\rho=   \sum_{i} \rho_{\bf i} \tilde{\sigma}_{\bf i} := \sum_{\bf i} \langle \tilde{\sigma}_{\bf i} , \rho  \rangle \tilde{\sigma}_{\bf i} ,
\end{equation}
and to estimate $\rho$ it suffices to estimate the Fourier coeffcients $\rho_{\bf i}$. 
A naive unbiased estimator can be easily constructed based on the counts of any particular measurement setting ${\bf d}$ 
for which $d_{j} = {\bf i}_{j}$ whenever ${\bf i}_{j}\neq 0$. For example when $k=2$, to estimate $ \rho_{(x,z)}$ we consider the counts from the setting ${\bf d}= (x,z)$, and define
$$
\hat{\rho}_{(x,z)}:= \frac{1}{\sqrt{2^{2}}n} \left[ 
N( (+1,+1 )|{\bf d}) +  N( (-1,-1 )|{\bf d}) - N( (+1,-1 )|{\bf d}) - 
N( (-1,+1)|{\bf d}) \right ] .
$$
While this proves that the state can be fully estimated, the naive estimator is generally not a bona-fide density matrix, and more importantly, has large estimation errors. The latter is due to the fact that $\hat{\rho}_{\bf i}$ is constructed from the counts of a \emph{single} setting and does not harness the information contained in the counts of the others. 
Indeed, since the projectors $\{ P_{\bf s}^{\bf d} :{\bf s}\in \mathcal{O}_{k} , {\bf d}\in \mathcal{D}_{k}\}$ form a (highly) overcomplete set of vectors in $M(\mathbb{C}^{2^{k}})$, any  product of Pauli's $\sigma_{\bf i}$ can be expressed in (continuously) many ways as a linear combination of projectors, each producing a linear estimator which could in principle be combined to obtain a significantly reduced MSE.  However, finding the ``optimal linear estimator'' is problematic due to the fact that the empirical frequencies $N^{\bf d}_{\bf s}/n$ are noisy estimates of the probabilities $\mathbb{P}({\bf s}|{\bf d})$, and their covariance depends on the unknown state. An interesting proposal in this direction is the Kalman filter estimator developed in \cite{Audenaert&Scheel}, but to our knowledge its performance in the case of MIT has not been extensively investigated. Another proposal put forward in \cite{Alquier&Butucea} is to combine the naive estimator with a second stage rank-penalised minimisation of the norm-two square (Hilbert-Schmidt) distance to the final estimator.

\emph{Maximum likelihood} (ML) is one of the most commonly used estimation methods across statistics. Its popularity is due to the intuitive interpretation, versatility, and strong theoretical underpinning.  Under certain regularity conditions the ML estimator is asymptotically optimal (or efficient in statistical terminology) in the sense that its covariance achieves the Cram\'{e}r-Rao bound in the limit of large samples, and has Normal (Gaussian) limiting distribution, with covariance equal to the inverse of the Fisher information matrix \cite{Young&Smith}. By discarding the constant factorial term in \eqref{eq.multinomial} and taking logarithm we can write the maximum likelihood estimator for MIT as
\begin{equation}\label{eq.ml}
\hat{\rho}:= \arg\max_{\tau\in \mathcal{S}_{k}}  \sum_{{\bf s},{\bf d}} N({\bf s}|{\bf d})  \log \mathbb{P}_{\tau} ({\bf s}|{\bf d}),
\end{equation}
where the maximum is computed over the set $\mathcal{S}_{k}$ of $k$-ions states $\tau$. Note that the ML estimator is invariant under reparametrisation, i.e. the ML estimator of a state functional $f:=f(\rho)$ is $f(\hat{\rho})$. The ML estimator has been used extensively in quantum statistics \cite{Hradil}, and an efficient iterative computational routine has been put forward in \cite{Hradil1997,Rehacek}. Nevertheless, ML has been criticised for several perceived drawbacks \cite{BlumeKohout,BlumeKohout2}. The first criticism is that the ML has the tendency to produce rank deficient estimators, when the true state has some small eigenvalues; this can be understood \cite{BlumeKohout} by  observing that the  likelihood (seen as a function of the matrix elements) may attain its maximum at a point which lies outside the convex space of states $\mathcal{S}_{k}$, in which case the ``constrained'' MLE $\hat{\rho}$ will fall on boundary of $\mathcal{S}_{k}$ by the concavity of the log-likelihood function. The second, and in our opinion more serious criticism is that the asymptotic theory does not apply as such, to states which lie on the boundary. As a side remark, we note that the asymptotic theory \emph{does} hold for the unconstrained ML estimator, for ``generic'' states which satisfy 
$\mathbb{P}_{\rho}({\bf s}|{\bf d})>0$ for all ${\bf s},{\bf d}$. This may be used to prove the existence of the asymptotic distribution of the MLE \eqref{eq.ml}, but the latter is likely to be complicated and impractical for establishing confidence regions (error bars). In this paper we focus on the performance of the proposed model selection estimation method, and refer to \cite{BlumeKohout3,Christandl&Renner} for two recent proposals for constructing confidence regions, and the forthcoming paper \cite{InnNott} for a comparative study of bootstrap and Fisher information methods. 


\section{Estimation of pure states in the MIT setting}
\label{sec.pure.states}

Pure states are arguably the golden standard of most state preparation experiments \cite{Haffner,Monz}. Therefore, we will start by considering the ideal situation in which the quantum state is assumed to be pure, and we would like to estimate it using the MIT dataset described in section \ref{sec.ion.tomography}. The goal is to get more insight into the statistical power of the measurement set-up and its asymptotic properties. This will prepare the ground for the next section where the purity assumption is lifted and the state is fitted to models of increasing rank, and model selection criteria are used for choosing a rank with a good trade-off between fit and model complexity. The findings are summarised at the end of the section, where we also clarify the relation to the compressed sensing set-up \cite{Gross&Liu&Flammia,Flammia&Gross} which uses as input the estimated values 
$\hat{\rho}_{\bf i}$ of the expectations of Pauli observables $\sigma_{\bf i}$.

\vspace{2mm}

The pure states ML estimator $\hat{\rho}$ can be computed as in \eqref{eq.ml}, with the maximisation restricted to the space of pure (rank one) states on $\mathbb{C}^{2^{k}}$. As figure of merit we consider the mean square error (MSE)
$$
MSE(\hat{\rho} ) := \mathbb{E}( \| |\rho - \hat{\rho} \|_{2}^{2}) 
$$
with the norm-two distance squared defined as 
\begin{equation}\label{eq.mse}
\|\rho - \hat{\rho}\|_{2}^{2} := \sum_{i,j=1}^{2^{k}} |\rho_{i,j} -\hat{\rho}_{i,j}|^{2}= \sum_{\bf i} |\rho_{\bf i}-\hat{\rho}_{\bf i}|^{2}. 
\end{equation}
where $\rho_{\bf i}$ are the Fourier coefficients with respect to the Pauli basis defined in \eqref{eq.pauli.fourier}. Note that for pure states the norm-two distance is related in a simple way to the (arguably more natural) norm-one distance $ \| \rho - \hat{\rho}\|_{1}:= {\rm Tr}(| \rho - \hat{\rho}|) $ by the equality 
$\|\rho - \hat{\rho} \|_{2} = \| \rho - \hat{\rho}\|_{1}/\sqrt{2}$.

\vspace{2mm}

We would like to address the following questions: 
\begin{itemize}
\item[1.] What is the MSE of the MLE  ?

\item[2.] Are we in an ``asymptotic regime'' ?

\item[3.] Is the multiple  ions measurement ``optimal'' in any sense ?

\end{itemize}

\vspace{2mm}

The pure states form a compact manifold of dimension $2(2^{k}-1)$ which can be identified with the complex projective space $\mathbb{C}P^{2^{k}+1}$. Therefore, when restricting the MIT statistical model to pure states, the standard asymptotic efficiency theory \cite{Young&Smith} is applicable. 
For simplicity, we assume that $|\psi\rangle$ has the expansion 
with respect to the standard basis $|\psi\rangle =\sum c_{i} |e_{i}\rangle$  such that $c_{1}\neq 0$, in which 
case we can parametrise the state  by the real and the imaginary parts of the remaining coefficients
$$
\theta\to |\psi_{\theta}\rangle = \sqrt{1-\|\theta\|^{2}} |e_{1}\rangle+ 
\sum_{j=2}^{2^{k}}  (\theta_{j} + i \theta_{2^{k} -2 +j }) |e_{j}\rangle, \qquad \theta\in \mathbb{R}^{2(2^{k}-1)} , \|\theta\|<1.
$$
Note that due to the geometry of the projective space, any global parametrisation must be singular unless some points are cut out as we did here. However, as we are interested in the asymptotic behaviour of the ML estimator, the global properties are unimportant and we can always choose an appropriate local parametrisation for all practical purposes. The norm two-square distance \eqref{eq.mse} can be rewritten locally as a quadratic form
\begin{equation}\label{eq.local.quadratic}
\|\rho_{\theta}- \rho_{\hat{\theta}}\|_{2}^{2} =  (\hat{\theta}- \theta)^{t} G(\theta) (\hat{\theta}- \theta)^{t} + o(\|\hat{\theta}- \theta\|^{2} ),
\end{equation}
where $G(\theta)$ is a positive definite matrix whose explicit form can be easily computed. The maximum likelihood estimator $\hat{\theta}= \hat{\theta}_{n}$ is efficient , i.e. as $n\to\infty$
\begin{equation}\label{eq.efficiency}
\sqrt{n}( \hat{\theta} -\theta) \overset{\mathcal{L}}{\longrightarrow} N(0, I(\theta)^{-1}), 
\end{equation}
where $N(0,I(\theta)^{-1}))$ is a the centered normal distribution with covariance matrix $I(\theta)^{-1}$ which is the inverse of the (classical) Fisher information matrix $I(\theta)$. In particular, from \eqref{eq.local.quadratic} and \eqref{eq.efficiency} we get
\begin{equation}\label{eq.trace.g.i}
\lim_{n\to\infty} \mathbb{E} ( \| \rho_{\theta}- \rho_{\hat{\theta}}\|_{2}^{2}) =  {\rm Tr} ( G(\theta)I^{-1}(\theta)).
\end{equation}

\begin{figure}[h]
\begin{center}
\includegraphics[width=8cm,height=7cm]{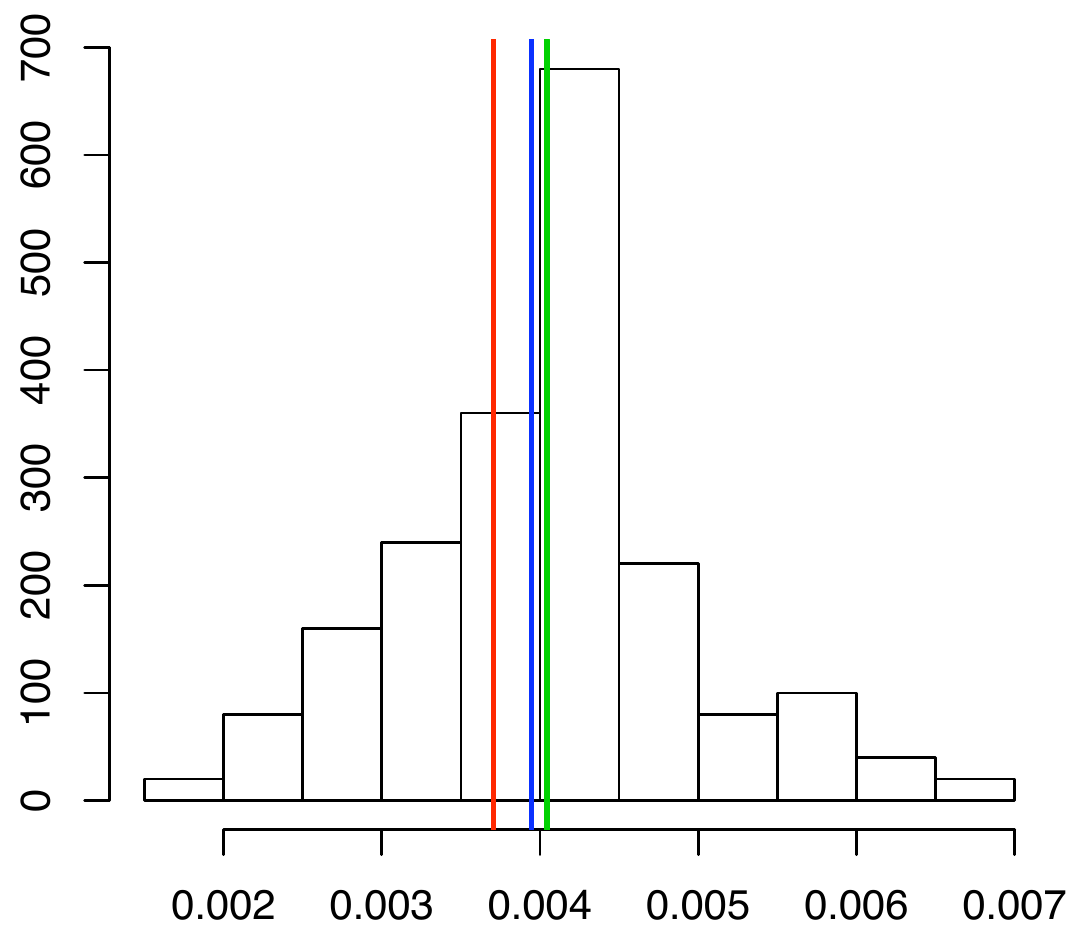}
\end{center}
\caption{Histogram of the norm-two error $\|\rho - \hat{\rho}\|^{2}_{2}$ of the MLE $\hat{\rho}$ for 100 samples from a fixed pure state $\rho$. The mean square error (green line) is very close to the classical Cram\'{e}r-Rao bound (blue line) as predicted by asymptotic theory, and the latter is only slightly larger than the  quantum Cram\'{e}r-Rao bound (red line), showing that for pure states the ions measurement is almost almost optimal among \emph{all} measurements. }\label{fig.pure.states}
\end{figure}

To verify these results we simulated 100 datasets from a fixed but randomly chosen pure state of $k=4$ ions, each 
dataset consisting of counts for $n=100$ measurements per setting. Figure \ref{fig.pure.states} shows the histogram of the square error $\|\rho_{\theta}- \rho_{\hat{\theta}}\|_{2}^{2}$ whose empirically estimated MSE (green line) is very close to the asymptotic prediction  (blue line) computed from
\eqref{eq.trace.g.i} which is equal to $3.9\cdot 10^{-3}$.  More interestingly, we find that the MSE is also very close to the 
``quantum optimal'' bound (red line) which describes the best MSE achievable with \emph{any} measurement! The latter is given by the simple formula ( see \cite{Gill&Guta} and references therein)
\begin{equation}\label{eq.qmse}
\text{QMSE}= \frac{\sharp \text{parameters}}{\sharp \text{samples}} = \frac{2(2^{k}-1)}{3^{k}\cdot n }=
\frac{2(2^{4}-1)}{3^{4}\cdot 100 }=  3.7\cdot 10^{-3}.
\end{equation}

This example shows that the MSE of the ML estimator agrees with the asymptotic theory when $n=100$ and $k=4$;  
we expect that the same holds for fixed $n$ and larger $k$ due to the favourable scaling of the number of samples  $n\cdot 3^{k}$ with respect to the number of parameters $2(2^{k}-1)$.  Moreover, the multiple ions measurement set-up appears to be quasi-optimal. This implies that adaptive strategies for choosing the settings cannot offer a significant improvement, but does not exclude the possibility that a similar performance can be achieved with a fraction of the settings. To further emphasise the point that the MIT dataset is \emph{very informative}, and that ML can optimally extract this information from the data, we compare the above results with the MSE of the naive estimator discussed in section \ref{sec.background}, and with the asymptotic MSE of a dataset consisting of \emph{estimates of the Pauli products} obtained by lumping together the counts of each measurement setting into a single statistic. 

\emph{MSE of the naive estimator.} With the square error defined as in \eqref{eq.mse},  we note  that the MSE of each coefficient $\hat{\rho}_{\bf i}$ for which $i_{1},\dots, i_{k}\neq 0$, is of the order 
$1/ (n\cdot 2^{k} )$ since we are essentially dealing with the problem of estimating the mean of a random variable with values $\{ +2^{-k/2} , -2^{-k/2} \}$. Therefore these coefficients alone (not counting those for which some $i_{j}$ are zero) 
bring a contribution of the order $3^{k}/(n\cdot 2^{k})$ which is larger than QMSE \eqref{eq.qmse} by a factor 
$(9/4)^{k}/2$. For the particular example of $k=4$ and $n=100$ this gives an MSE of $5\cdot 10^{-2}$ which is an order of magnitude larger than that of the MLE.

\begin{figure}[h]
\begin{center}
\includegraphics[width=6cm,height=6cm]{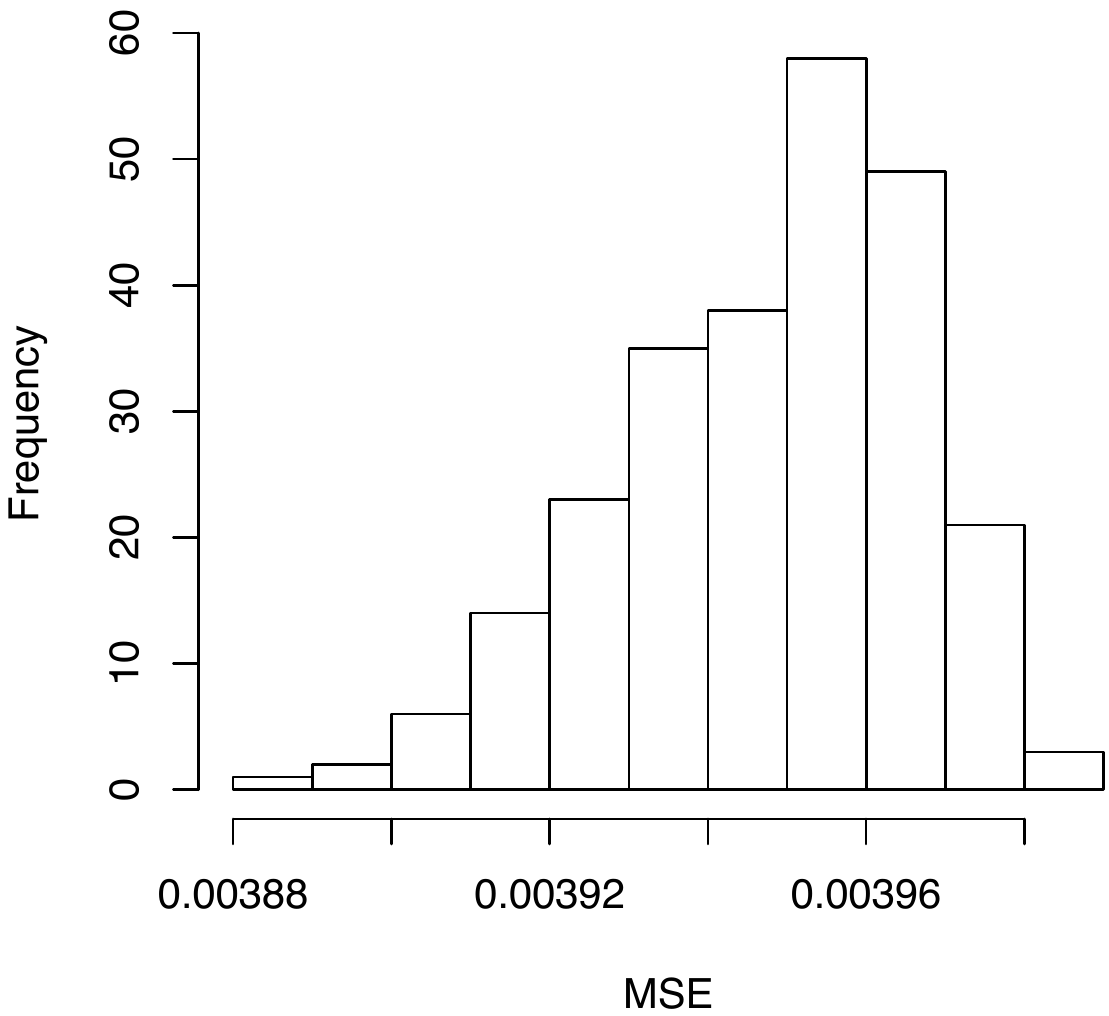}\hspace{4mm}
\includegraphics[width=6cm,height=6cm]{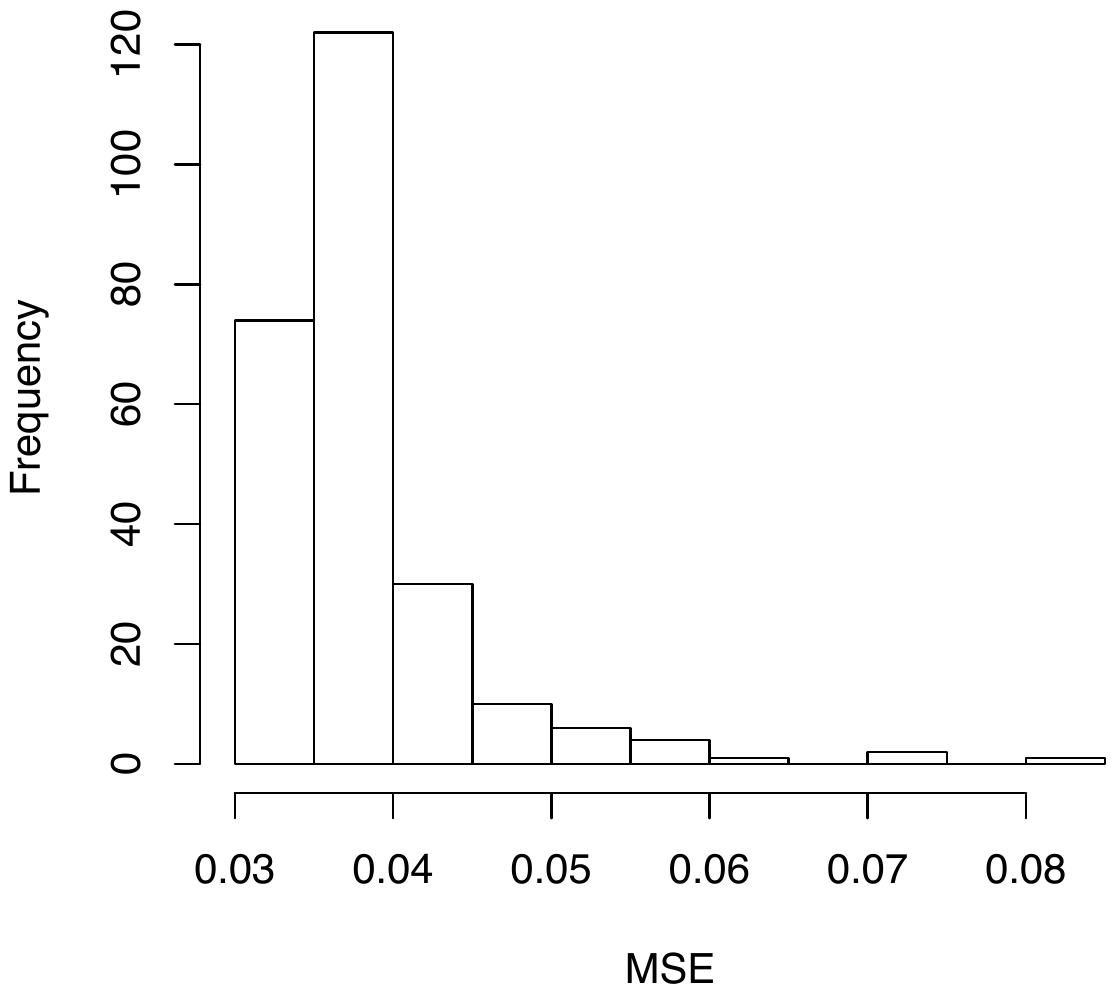}
\end{center}
\caption{Histograms of asymptotic MSE's for 250 randomly chosen pure states, with $k=4$ and $n=100$. Left panel: full counts dataset. Right panel: coarse grained dataset. Keeping only the empirical means of the Pauli products leads to a 10 fold increase in the MSE.}
\label{fig.compressed.sensing}
\end{figure}

\emph{MSE of the coarse grained data.} At this point, it is natural to ask the following question. Suppose that we are given the $3^{k}$ empirical averages of  the Pauli products $\sigma_{\bf i}$
\begin{equation}\label{eq.lumped.estimators}
\hat{\rho}_{\bf i} \approx  {\rm Tr}(\rho \tilde{\sigma}_{\bf i} ) = \langle \psi|  \tilde{\sigma}_{\bf i} |\psi\rangle, 
\qquad i_{1},\dots i_{k}\neq 0
\end{equation}
which are obtained by computing one empirical average for each column of the original dataset. 
Is there a more efficient method to estimate the pure state $|\psi\rangle$, from the data \eqref{eq.lumped.estimators} 
and what is its MSE ? Two important candidates are the \emph{compressed sensing} and \emph{lasso} algorithms 
\cite{Flammia&Gross} (with the slight difference that they would use a smaller number of settings, but proportionally more measurements per setting). Both methods aims at estimating the state by trying to match the empirical expectations 
$\hat{\rho}_{\bf i}$ with those of a selfajoint matrix, while in the same time penalising the trace norm of the matrix. Testing these methods is beyond the scope of this paper, but the asymptotic efficiency theory offers a shortcut to the answer of the above question. Applying the same methodology as before, but to the coarse grained data \eqref{eq.lumped.estimators} we can predict that (asymptotically) the MSE of any estimator is bounded from below by that of the ML estimator $\hat{\rho}_{\rm cg}$ which in turn satisfies
\begin{equation}\label{eq.lumped.mse}
\lim_{n\to\infty} n \mathbb{E}\left( \| \hat{\rho}_{\rm cg} - \rho\|_{2}^{2}\right)  = {\rm Tr}(G(\theta) I_{cg}(\theta)^{-1})
\end{equation}
where the only difference with \eqref{eq.qmse} is the Fisher information matrix which satisfies the inequality $I_{cg}(\theta)  \leq I(\theta)$. Figure \ref{fig.compressed.sensing} shows histograms of the asymptotic MSE \eqref{eq.qmse} for the full MIT data (left panel) versus the MSE \eqref{eq.lumped.mse} of the coarse grained data (right panel). The histograms were produced with 250 randomly chosen pure states, $k=4$ and $n=100$.  
Note that the MSE of the coarse grained data is smaller than the (partial) estimated contribution of the naive estimator. However, the MSE is still an order of magnitude higher that that of the full dataset, due to the fact that a significant amount of information has been discarded in the process of retaining the Pauli products expectations.

To summarise, we conclude that MIT works because the different settings ``overlap'' with each other in the sense that the one dimensional projections $|e_{\bf d}^{\bf s}\rangle\langle e_{\bf d}^{\bf s}| $ form an overcomplete set of size $2^{k}\times 3^{k}$ which is significantly larger than the dimension of the space of matrices $4^{k}$ even for small $k$. Therefore, the measurement data is structured so that the counts for each setting provide a relatively small amount of information, but the dataset as a whole is very informative about the state. Reducing this dataset to a small number of expectations may be  advantageous for the purpose of devising fast estimation algorithms, but underperforms from the viewpoint of statistical errors, for a given number of state re-preparations. This statement may seem to contradict the simulation results illustrated in Figure 1 of \cite{Flammia&Gross}, where compressed sensing and lasso are found to perform \emph{better} than ML on datasets of the type \eqref{eq.lumped.estimators}. This apparent contradiction is lifted by the following observations:

\begin{itemize}
\item[1)] 
Our comparison is between the MSE of efficient estimators for two \emph{different types of data}. 
Based on this we conclude that 
\emph{any} estimator using the coarse grained data will asymptotically underperform ML based on the full counts experimental dataset. 

\item[2)]
The comparison in \cite{Flammia&Gross} is different; it regards the performance of ML versus compressed sensing and lasso for the coarse grained data. Since a completely unknown state is \emph{not identifiable} for the coarse grained model, the MLE is \emph{not consistent}, and arguably should not be used in this case. 

\end{itemize}

\section{Model selection for quantum tomography}
\label{sec.model.selection}

In the previous sections we discussed the extreme scenarios of ``full'' quantum tomography and  estimation of pure states. In reality, the states produced in experiments tend to have one or few significant eigenvalues and a large number of small eigenvalues of different orders of magnitude, which account for the imperfections in the preparation procedure. Therefore, neither of the two settings seems to be suitable: the former underfits the real state while the latter overfits by trying to estimate eigenvalues that may not be statistically significant. 

This is a well known phenomenon in statistics, that occurrs in high (or infinite) dimensional problems such as estimating the probability density of a real valued random variable, or in non-linear regression where an unknown function is ``learned'' from estimates of its values at certain points. In such cases the maximum likelihood estimator overfits the data and is very ``noisy''. A possible solution is to use a \emph{penalised maximum likelihood} estimator, which maximises the difference between the log-likelihood and a penalty measuring the complexity of the estimator, e.g. the number of non-zero  coefficients with respect to an appropriate basis. 
More generally, one can design a class of statistical models with various degrees of complexity, and decide which 
model and which estimator from that model is most suitable for describing the data. Our aim is to apply the model selection methodology to state tomography, the models being the families of states of a given rank. The same methods can be used in tasks such as quantum homodyne tomography \cite{Vogel&Risken} where the state to be estimated is that of a light pulse (one mode continuous variables system), and a model could be the set of states with a given maximum number of photons \cite{Guta&Artiles}. 

To select the rank of the state we will use two well established methods: the Akaike information criterion (AIC) \cite{Akaike} and the Bayesian information criterion (BIC) \cite{Schwarz}. Both methods amount to penalising the log-likelohood function 
by a factor proportional to the dimension of the model, and choosing the ML estimator with the smallest value of the information criterion. In the next section we give a brief general description of AIC and BIC, after which we discuss the parametrisation of the fixed rank quantum models, and the implementation of the model selection procedure.


\subsection{AIC versus BIC model selection}


Occam's razor is an old scientific principle which states that when trying to explain a phenomenon, one should choose 
\emph{the simplest} model that adequately fits the data. 
A very complex model will be able to fit the given data almost perfectly but it will not be able to generalise very well. On the other hand, very simple models will not be able to describe the essential features of the data. Therefore, we must make a compromise and choose a model which is as simple as possible, but no simpler. It is not surprising that many approaches have been proposed over the years for dealing with this key aspect in statistical modelling.

The general framework of model selection is the following. We are given $n$ samples ${\bf X}= \{X_{1},\dots , X_{n}\}$ from some unknown distribution $\mathbb{P}$ which we try to fit with a distribution from one of several possible statistical models 
$$
\mathcal{M}_{r} := \{ \mathbb{P}_{\theta_{r}} :\theta_{r}\in \Theta_{r}\subset \mathbb{R}^{p(r) } \}, \qquad r=1,\dots ,D,
$$
where  $\mathcal{M}_{r}$ has a parameter $\theta_{r}$ of dimension $p(r)$. For simplicity we assume that 
$X_{i}\in \{1,\dots, a\}$ are discrete random variables, as in the case of MIT measurements.
We also assume that at least one of the $D$ models contains the true distribution, or at least gives a reasonabale approximation to it. We denote by $\hat{\theta}_{r}$ the ML estimator for the model $\mathcal{M}_{r}$, and by 
$\ell_{\theta_r}= \ell_{\theta_r}({\bf X}):= \sum_{i=1}^{n} \log \mathbb{P}_{\theta_r}(X_{i})$ log-likelihood function at $\theta_{r}$.


\subsubsection{Akaike Information Criterion (AIC)}


The AIC for model $\mathcal{M}_{r}$ is \cite{Akaike}
$$
\mbox{AIC}(r) = - 2 \cdot \ell_{\widehat{\theta}_r}+ 2p(r),
$$ 
and the chosen model is the one with the minimum AIC. Since $p(r)$ is larger for more complex models, the AIC formally biases against overly complicated models. Although the derivation of AIC is outside the scope of this paper, we briefly explain the idea behind the choice of penalty. Having computed the ML estimators for different models we would like to select the ``best'' one in the sense that the corresponding distribution $\mathbb{P}_{\hat{\theta}_{r}} $ is the closest to the ``truth '' $\mathbb{P}$ with respect to the Kullback-Leibler distance (or relative entropy) 
$$
K( \mathbb{P}|\mathbb{P}_{\hat{\theta}_{r}}) := \sum_{i=1}^{a}  \mathbb{P}(i)  \log( \mathbb{P}(i)) -  
\sum_{i=1}^{a}  \mathbb{P}(i) \log( \mathbb{P}_{\hat{\theta}_{r}}(i) ).
$$
However, this quantity cannot be computed since $\mathbb{P}$ is unknown. Since the first terms on the right side is the same for all models, it can be neglected, and one can focus on estimating the second term, 
which nevertheless still depends on $\mathbb{P}$. If instead of $\hat{\theta}_{r}$ we had a fixed parameter $\theta_{r}$, this term would be the expected  value of the log-likelihood at $\theta_{r}$ and could be estimated by $\ell_{\theta_{r}}({\bf X})/n$, by the law of large numbers. However $\ell_{\hat{\theta}_{r}}({\bf X})/n$  is a biased estimator of the second term, due to the fact that the data has been already used in computing $\hat{\theta}_{r}$. Akaike showed that under the regularity conditions required by the asymptotic normality theory,  the bias is approximately $p(r)/n$, so that ML which is the closest to the truth is approximately give by the minimizer of the AIC.

\subsubsection{Bayesian Information Criterion (BIC)}
The BIC for model $\mathcal{M}_{r}$ is defined as \cite{Schwarz}
$$
\mbox{BIC}(r) = - 2 \cdot \ell_{\hat{\theta}_r} + p(r) \log{(n)}
$$ 
where $n$ is the sample size.  Note that the BIC differs from the AIC only in the second term which increases with $n$, so that BIC favors simpler models (that is models with a smaller number of parameters) compared to AIC. But despite the superficial similarity between the AIC and BIC the latter is derived in a very different way, within a Bayesian framework.

For simplicity, suppose that there are two competing models, $\mathcal{M}_1$ and $\mathcal{M}_2$ with parameters 
$\theta_1$ and $\theta_2$ respectively. One begins by assigning prior probabilities $q_{1}$ and $q_{2}=1-q_{1}$  to the event that the observed data have been generated from either model. One also assigns prior distributions  $\pi_{1}(\theta_{1})$  and $\pi_{2}(\theta_{2})$ to the model parameters in each model. Then one can compute the marginal likelihoods which can be interpreted the probability of observing the data if model $\mathcal{M}_i$ is correct, having integrated out our ignorance about the parameters $\theta_1$ and $\theta_2$ in each model. Hence, one can apply Bayes theorem to evaluate the probability of model $\mathcal{M}_i $ being the true model given the observed data. A measure of the extent to which the data support model $\mathcal{M}_2$ over $\mathcal{M}_1$ is given by the \emph{posterior odds} 
$$
\frac{\mathbb{P}(\mathcal{M}_2| {\bf X} )}{\mathbb{P}(\mathcal{M}_1|{\bf X})} 
= \frac{\mathbb{P}({\bf X} | \mathcal{M}_2)}{ \mathbb{P}{\bf X} | \mathcal{M}_1)} \frac{q_{2}}{q_{1}}
$$
The first fraction on the right-hand side is called the \emph{Bayes factor} and the second is known as the prior odds. The Bayes factor is a fundamental quantity in Bayesian theory and can be interpreted as a measure of the extent to which the data support model $\mathcal{M}_2$ over $\mathcal{M}_1$ when the prior odds are equal to one. The difference $BIC(1)- BIC(2)$ can be shown to be a large sample approximation to the logarithm of the Bayes factor, so that the second model is chosen if the difference is positive.

\subsection{Parametrising models with fixed rank}

Here we describe the fixed rank models which will be used in model selection. Let $\mathcal{D}(d,r)$ be the set of rank $r$ states of a $d$-dimensional quantum system, i.e. those states which have exactly $r$ non-zero eigenvalues, and let
$$
\mathcal{R}(d,r):= \bigcup_{i=1}^{r} \mathcal{D}(d,r)
$$
be the set of states of rank at most $r$.
Every state $\rho$ has a unique spectral decomposition
$$
\rho= \sum_{i=1}^{r} \lambda_{i} P_{i} 
$$
where $\lambda_{i}>0$ are its \emph{distinct} eigenvalues, and $P_{i}$ is an eigenprojector whose dimension is equal to the multiplicity $m_{i}$ of $\lambda_{i}$.  The spectral information $(\lambda_{1}, P_{1} , \dots, \lambda_{r} , P_{r})$ can be used to construct a parametrisation of $\mathcal{D}(d,r)$ and $\mathcal{R}(d,r)$, which has the advantage of a direct physical interpretation. However, the practical implementation of such a parametrisation for computing the maximum likelihood estimator is less straightforward due to the orthogonality constraints for the eigenvectors, and the singularities appearing on lower dimensional manifolds consisting of states with non-trivial sets of multiplicities. A variation on this would  be to parametrise the state by the set of eigenvalues and an eigenbasis, in which case the singularity problem is replaced by the 
\emph{non-identifiability} of the different basis vectors corresponding to the same eigenvalue.

We will describe an alternative parametrisation which is related to the Cholesky factorisation of the state. Recall that any \emph{positive definite} matrix 
$A\in M(\mathbb{C}^{d}) $ has a unique decomposition
\begin{equation}\label{eq.cholesky}
A= T^{*}T
\end{equation}
where $T$ is an upper triangular matrix with strictly positive diagonal elements. Therefore there exists a one-to-one correspondence between full-rank states $\rho$ and matrices $T$ as described above, with the additional constraint 
\begin{equation}\label{eq.normalisation.t}
{\rm Tr} (T^{*}T) = \sum_{ij}|T_{ij}|^{2} = {\rm Tr}(\rho) =1.
\end{equation}
We parametrise such a matrix $T$ by the vector of real numbers 
$\theta:= (R, I, D)\in \mathbb{R}^{d^{2}-1}$ with
\begin{equation}
\left\{
\begin{array}{lll}
R&:= & \left({\rm Re}(T_{12}) , \dots,  {\rm Re}(T_{d-1,d})\right) \\
I  &:= &\left({\rm Im}(T_{12}),\dots , {\rm Im}(T_{d-1,d})\right)\\
D &:= &(T_{22},\dots,T_{dd} )
\end{array}
\right.
\label{eq.parametrisation}
\end{equation}
such that $R,I$ are the real and imaginary parts of the off-diagonal elements ordered from the first to the $d-1$ row, 
and from left to right along each row. By \eqref{eq.cholesky} and \eqref{eq.normalisation.t},  $\theta$ must satisfy the constraints $D>0$ and $ \| R\|^{2} + \| I \|^{2}+ \| D \|^{2}<1$, and the left-top element of $T$ is equal to
$$
T_{11}= T_{11}(\theta) =(1 - \|R\|^{2} + \|I\|^{2}+ \|D\|^{2} )^{1/2} >0.
$$

The  Cholesky parametrisation of the full rank matrices can be extended, albeit with some caveats, to the spaces of rank-deficient matrices $\mathcal{D}(d,r)$ and $\mathcal{R}(d,r)$. The idea is to consider a decomposition as in \eqref{eq.cholesky}, but with $T$ belonging to the set $\mathcal{T}^{+}(d,r)$ of $d\times d$ upper triangular matrices with the bottom $d-r$ rows equal to zero, and satisying $T_{11},\dots ,T_{rr}>0$; equivalently, one can consider  $r\times d$ trapezoidal matrices obtained by removing the zero lines of the triagular matrices. Since every $T\in \mathcal{T}^{+}(d, r)$ is of rank $r$, this guarantees that the corresponding state $\rho$ has the same property. However, not all states of rank $r$ can be decomposed in this way! Indeed it is easy to verify that if $\rho= T^{*}T$ then the $r\times r$ top-left principal minor of $\rho$ must be of rank $k$, and therefore such a parametrisation excludes states in $\mathcal{D}(d,r)$ which do not satisfy this property. Nevertheless,  ``generic'' matrices of rank $r$ \emph{do} have principal minors  of rank $r$, in the sense that those with smaller rank principal minor form a lower dimensional subset of $\mathcal{D}(d,r)$. If restrict our attention to the subset $\mathcal{D}(d,r)^{+}\subset\mathcal{D}(d,r)$ which excludes the ``deficient'' states, we find that the Cholesky decomposition exists and is unique, so that 
$$
\mathcal{D}(d,r)^{+}:= \{ \rho =T^{*}T : T\in  \mathcal{T}^{+}_{d,r} \} \subset \mathcal{D}(d,r).
$$

What can we say about the complement $\mathcal{D}(d,r)\setminus \mathcal{D}(d,r)^{+}$ ? In order to have a Cholesky decomposition we need to relax the condition  $T_{11},\dots ,T_{rr}>0$ and consider the set   $\mathcal{T}(d,r)$ of $r$-lines upper triangular matrices, with non-negative elements on the diagonal. In this case, the root  $T$ not only exists but is in general not unique.

Let $\Theta(d,r)^{+} $ be the set of real parameters $\theta:= (R,I,D)$ of a matrix 
$T= T_{\theta}\in\mathcal{T}(d,r)^{+}$ which are defined similarly to equation \eqref{eq.parametrisation}, and let  $\Theta(d,r)$ be the set of parameters associated to matrices in $\mathcal{D}(k,r)$. We define two \emph{sequences} of quantum statistical models:
\begin{eqnarray}
\mathcal{Q}^{+} (d,r)& := &\{ \rho_{\theta}= T_{\theta}^{*}T_{\theta} : \theta \in \Theta(d,r)^{+} \} , \qquad r=1,\dots , d\\
\mathcal{Q} (d,r)& := &\{ \rho_{\theta}= T_{\theta}^{*}T_{\theta} : \theta \in \Theta(d,r)\}  , \qquad r=1,\dots , d
\end{eqnarray}
the first one consisting of rank $r$ matrices with rank $r$ principal minor, the second one describing (albeit not always uniquely) all matrices of rank up to $r$. The reason why we mention the two models is that each has some appealing features and some disadvantages. For $\mathcal{Q} (d,r)$ the advantage is that we deal with a \emph{nested} set of models
$$
\mathcal{Q} (d,1)\subset \mathcal{Q} (d,2)\subset\dots \subset \mathcal{Q} (d,d).
$$ 
 The disadvantage is that the Cholesky parametrisation is not one-to-one in this case. On the other hand, 
 $\mathcal{Q} (r,d)^{+}$ offers a one-to-one parametrisation of  rank $r$ matrices in  $\mathcal{D}(d,r)^{+}$, with the disadvantage that the models are not nested, but instead $\mathcal{Q}(d,r)^{+}$ lies on the boundary of $\mathcal{Q}(d,r+1)^{+}$. While these facts are relevant to a theoretical analysis, for practical purposes the distinction between the two models is less important, and in all our numerical experiments we used the models $\mathcal{Q}^{+} (d,r)$.

\subsection{The implementation of AIC and BIC model selection for rank-based models}

We return now to the state estimation problem, and describe how AIC and BIC model selection is applied to the family of rank-based models described above for a system consisting of $k$ ions, i.e. $d= 2^{k}$. Let 
$$
\ell_{\theta} =\ell_{\theta} \left( \{  N({\bf s}|{\bf d})  : {\bf s} \in \mathcal{O}_{k}, {\bf d} \in\mathcal{D}_{k} \}\right) 
:= \sum_{{\bf s},{\bf d}} N({\bf s}|{\bf d})  \log \mathbb{P}_{\rho_{\theta}} ({\bf s}|{\bf d})
$$
be the log-likelihood of the measurement data, ignoring the constant factorial terms. The maximum likelihood estimators 
$\hat{\theta}_{r} $ and $\hat{\rho}_{r}$ for the model 
$\mathcal{Q}(2^{k},r)^{+}$ are :
$$
\hat{\theta}_{r} :=  \arg\max_{\theta\in \Theta(2^{k},r)^{+}} 
\ell_{\theta}, 
\qquad
\hat{\rho}_{r}:= \rho_{\hat{\theta}_{r} }.
$$
In order to choose between the different models we compute the AIC and the BIC for each rank and select the model with the smallest value. In our case the two criteria are given by
\begin{equation}
\left\{
\begin{array}{cccc}
AIC(r)&:= -2 \ell_{\hat{\theta}_{r}} +&  2 p(2^{k},r)&, \\[2mm]
BIC(r)&:=-2 \ell_{\hat{\theta}_{r}}  + & p(2^{k},r) \log (n\cdot 3^{k})&,
\end{array}
\right.
\end{equation}
with 
\begin{equation}\label{eq.dimension.model}
p(d,r)= 2\cdot d\cdot r-r^{2}-1
\end{equation} 
the dimension of the space of rank $r$ matrices, and $n\cdot 3^{k}$ is the total number of measurements. In practice each  criterion decreases with the rank until it reaches the minimum value after which it increases,  so one only needs to compute the ML estimator up to the rank where the criterion begins to increase. For low rank states, this offers a the advantage of having to compute the maximum likelihood estimator on models of dimension approximately $r\cdot d$ rather than $d^{2}-1$ as standard ML. The disadvantage is that the likelihood function is not concave as in the full rank model, and may have several local maxima. 

To implement the ML estimation numerically, we used a standard maximisation routine of the statistics package {\tt R}.  Additionally, we developed an array of statistical analysis tools such as Fisher information, square errors, bootstrap, Pearson $\chi^{2}$ statistic which will be made available online. Although the computation of the log-likelihood was optimised for faster speed, the maximisation can probably be improved significantly by using more sophisticated routines.

In the next sections we will discuss the results of several investigations on the performance of BIC and AIC model selection, using simulated and real data.

\section{Study 1: randomly chosen low rank states} 
\label{sec.low.rank.states}

In a first simulation study we chose 3 ``random'' states of ranks 1, 2,  and 3 of $k=4$ ions, and generated 100 datasets from each state, each dataset with $n=100$ measurement repetitions. We then computed the maximum likelihood estimators for the ranks between 1 and 4 and used AIC and BIC to select the optimal rank. The exact procedure used to generate ``random'' states is not very important, but it will be relevant that all non-zero eigenvalues of the states are significant. As illustrated in Table \ref{table.BIC.random.state}, BIC selected the correct rank for each state in roughly 90\% of the cases while for AIC the rate is about 80 \%.  Due to the different penalties, the AIC tends to over-estimate the rank of the state, while BIC has a slight tendency to under-restimate it. 
\begin{table}[h]
\centering
\begin{tabular}{ c | c | c | c | c| c }
\cline{2-5}
& \multicolumn{4}{| c |}{AIC rank} \\
\cline{1-5}
\hline
\multicolumn{1}{| c |}{ true rank} \vline & 1 & 2 & 3 & 4 \\
\cline{1-5}
\multicolumn{1}{| c |}{1}  \vline & 82 & 9 &9 &0 \\
\cline{1-5}
\multicolumn{1}{| c |}{2}  \vline & 0& 74 & 26 & 0 \\
\cline{1-5}
\multicolumn{1}{| c |}{3}  \vline & 0& 1 & 80 & 19\\
\cline{1-5}
\end{tabular}
\hspace{10mm}
\begin{tabular}{ c | c | c | c | c| c | c | c | c}
\cline{2-5}
& \multicolumn{4}{| c |}{BIC rank} \\
\cline{1-5}
\hline
\multicolumn{1}{| c |}{ true rank} \vline & 1 & 2 & 3 & 4 \\
\cline{1-5}
\multicolumn{1}{| c |}{1}  \vline & 99 & 0 &1 &0 \\
\cline{1-5}
\multicolumn{1}{| c |}{2}  \vline & 7& 90 & 3 & 0 \\
\cline{1-5}
\multicolumn{1}{| c |}{3}  \vline & 0& 5 & 95 & 0\\
\cline{1-5}
\end{tabular}
\caption{AIC and BIC performance for 100 datasets generated by 3 randomly chosen states of ranks 1, 2 and 3. 
The tables shows the number of times AIC and BIC choose rank 1, 2 or 3 for each state.}
\label{table.BIC.random.state}
\end{table}
While at first sight this may appear to be a surprisingly good performance, we will show that it agrees very well with the predictions of asymptotic theory. For illustration, we consider the state of rank $r=2$ denoted $\rho$, and show that the distributions of  $BIC(3)- BIC(2)$ and $BIC(1)- BIC(2)$ concentrate on the positive axis, so that BIC chooses the correct rank. Since their behaviours are determined by different mechanisms, we will study each BIC difference separately. A similar analysis can be performed for AIC.

\begin{figure}[h]
\begin{center}
\includegraphics[width=6.2cm]{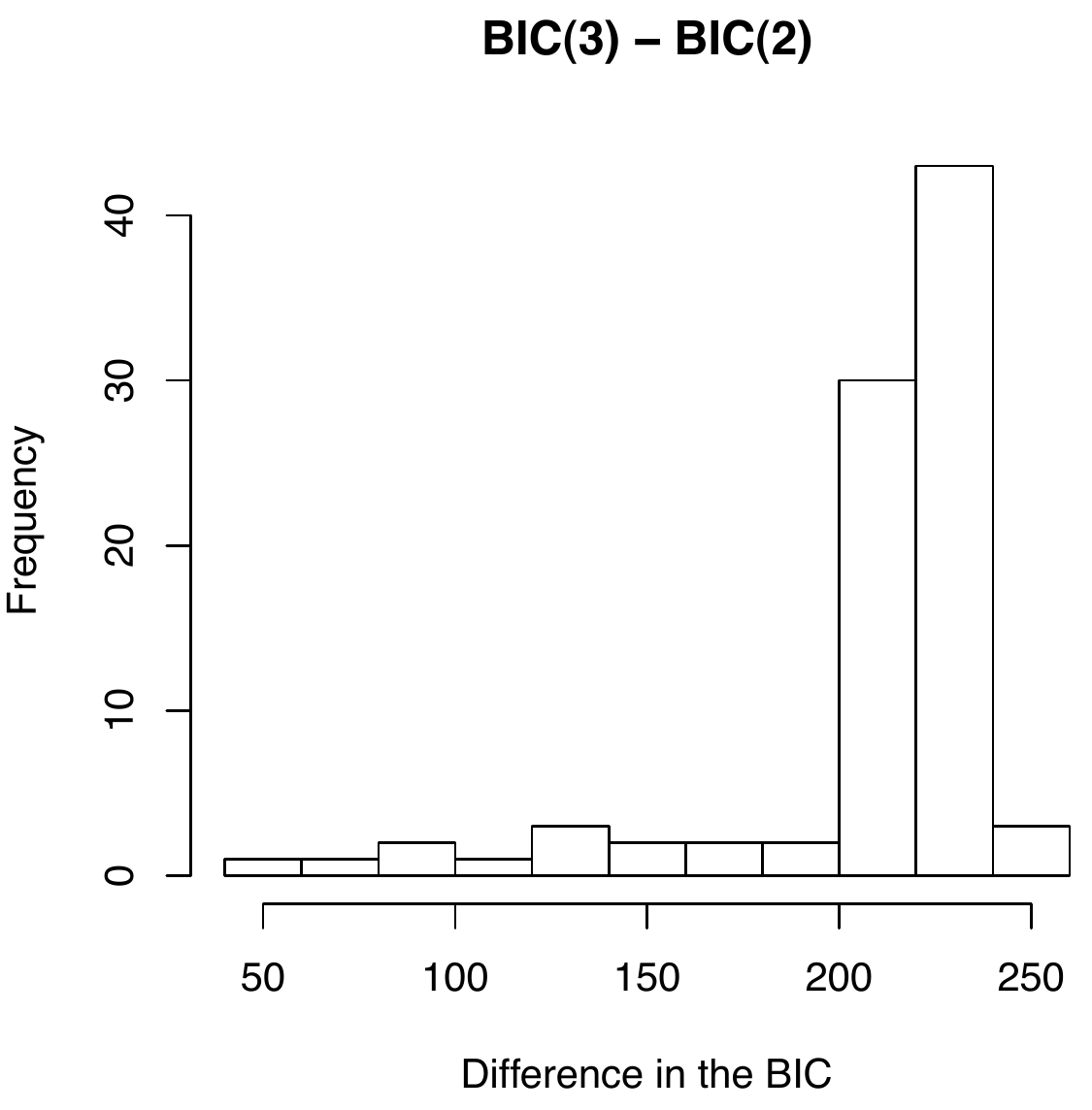}\hspace{1mm}
\includegraphics[width=6.2cm]{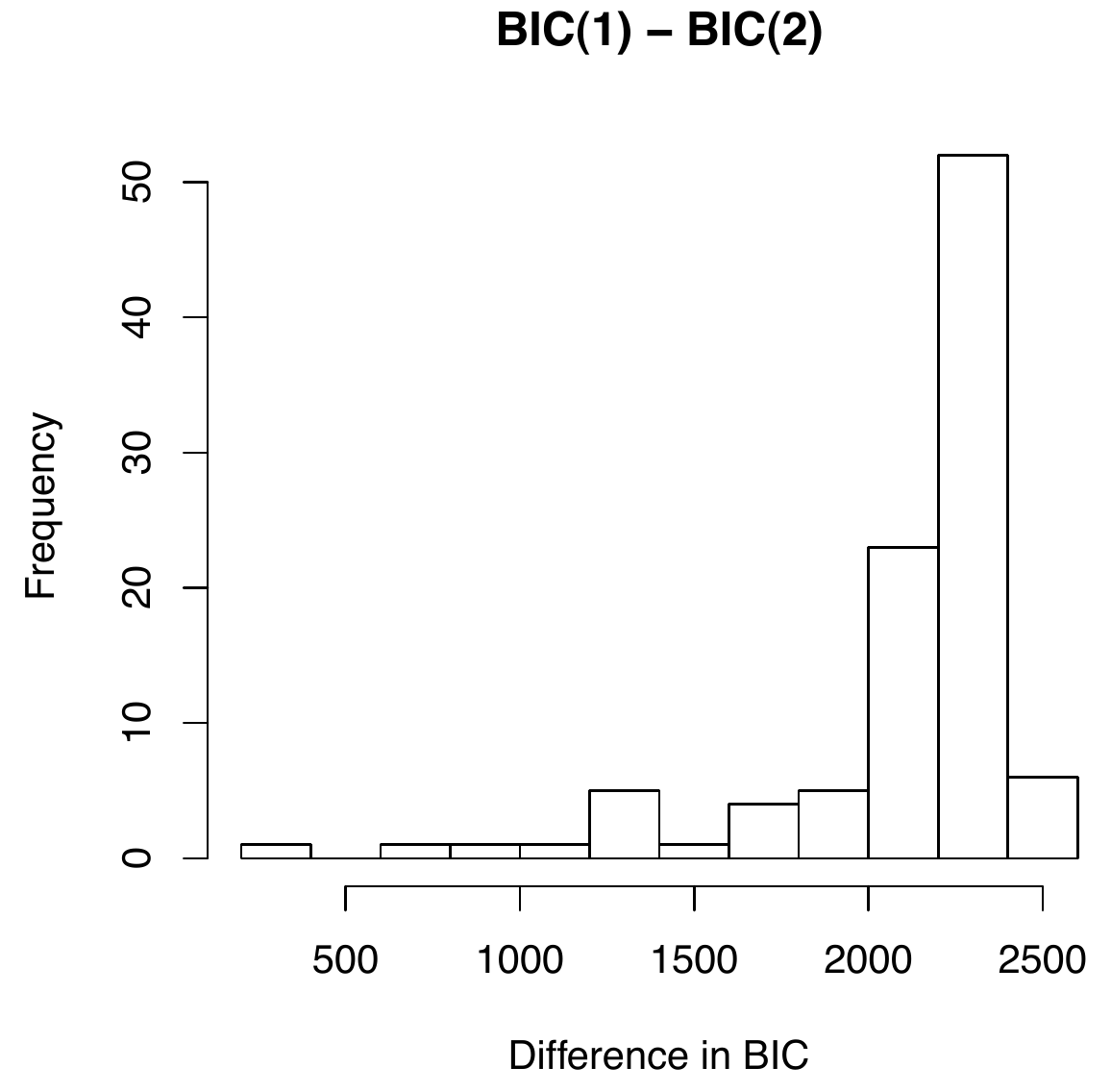}
\end{center}
\caption{Histogram of BIC differences for the rank 2 state. Left panel $BIC(3)- BIC(2)$; right panel 
$BIC(1)- BIC(2)$. The values are in good agreement with the asymptotic predictions.}\label{fig.BIC-rank}
\label{fig.diff.bic}
\end{figure}

In the first case,
\begin{eqnarray}
BIC(3)- BIC(2) &=& 
-2( \ell_{\hat{\theta}_{3} } -  \ell_{\hat{\theta}_{2} }) + \log (n \cdot 3^{4})(  p(4,3) - p(4,2) )\nonumber
\\
&=& -2(\ell_{\hat{\theta}_{3} } -  \ell_{\hat{\theta}_{2} }) + 242.98
\label{eq.bic.difference}
\end{eqnarray}
so the problem is to show that the \emph{log-likelihood ratio statistic} 
$$
\Lambda:= 2(\ell_{\hat{\theta}_{3} } -  \ell_{\hat{\theta}_{2} }).
$$  
is typically smaller than the penalty $ 242.98$. For ``regular'' nested models, the asymptotic distribution of $\Lambda$ is described by  Wilks' theorem as discussed in section \ref{sec.stats}. However, this is not directly applicable here since the rank 2 model lies on the boundary of the rank 3 one, due to the positivity constraints. Nevertheless, Wilks' theorem can be extended to more general situations where the two hypotheses can be ``linearised'' locally (see \cite{vanderVaart} chapter 16), in which case the limiting distribution depends on the local geometry of the two models and the Fisher information at each point. We will not pursue this analysis here but limit ourselves to giving a stochastic upper bound to the limiting distribution which will be sufficient for our purposes. The idea is to note that 
\begin{equation}\label{loglik.ratio.inequality}
\Lambda:= 2(\ell_{\hat{\theta}_{3} } -  \ell_{\hat{\theta}_{2} }) \leq  2(\ell_{ \tilde{\theta}_{3}}  -  \ell_{\hat{\theta}_{2} })
\end{equation}
where $\ell_{\tilde{\theta}_{r}}$ is the ``unconstrained'' maximum likelihood estimator obtained by maximising over the  space $\tilde{\mathcal{D}}(d,r)\supset \mathcal{D}(d,r)$ consisting of matrices $\rho$ of rank $r$ which are not necessarily positive but must respect the property that $\mathbb{P}_{\rho}(\cdot |{\bf d})$ is a probability distribution for each ${\bf d}$. The unconstrained MLE is easier to analyse theoretically and can be used to explain why MLE often produces rank deficient states when the true state has high purity \cite{BlumeKohout}. Now, assuming that we are in the generic situation where all probabilities for the true rank-two state $\rho$ are non-zero, this means that locally around $\rho$ the rank two model is a regular submodel of the extended rank 3  model, and we can apply Wilks' theorem to conclude that 
$$
2(\ell_{\tilde{\theta}_{3}} -\ell_{\hat{\theta}_{2}}) \overset{\mathcal{L}}{\longrightarrow} \chi^{2}(p(4,3) - p(4,2) ).
$$
From \eqref{loglik.ratio.inequality} we get that $\Lambda$ is stochastically bounded from above by $\chi^{2}(27)$ and similarly $BIC(3)- BIC(2) $ is bounded from below by $242.98 -\chi^{2}(27)$ which agrees with the simulations results illustrated in the left panel of Figure \ref{fig.diff.bic}. Note that as $n$ increases, the probability of BIC choosing the rank 3 model converges to zero due to the presence of the $\log n$ factor in the penalty, while AIC is \emph{not rank consistent} in the sense that it choses the higher rank with a probability which does not vanish with $n$, in agreement with the results illustrated in Table \ref{table.BIC.random.state}.
Let us consider now the second difference $BIC(1)- BIC(2)$, and note that its values are much larger, a illustrated in the right panel of Figure \ref{fig.diff.bic}. It turns out that in this case the behaviour is not dominated by the complexity 
penalty but by the \emph{bias} of the lower rank model with respect to the ``correct'' one, and in particular the distribution of the difference is state dependent. The key is to observe that while the rank 2 ML estimator $\hat{\rho}_{2}$ converges to the true state $\rho$, the rank one ML estimator $\hat{\rho}_{1}$ converges to the state $\rho^{*}_{1}$ whose corresponding distribution  $\mathbb{P}_{\rho_{1}^{*}}$ is the closest to the true distribution $\mathbb{P}_{\rho}$ with respect to the  relative entropy (or Kullback-Leibler divergence) 
$$
\rho_{1}^{*}:= \underset{\tau\in \mathcal{D}(2^{4}, 1)}{\arg\min} ~
K\left( \mathbb{P}_{\rho} \left| ~\mathbb{P}_{\tau}  \right.\right).
$$
In conjunction with the law of large numbers we then obtain the almost sure convergence
$$
\frac{\Lambda}{2n} = \frac{1}{n}\left( \ell_{\hat{\theta}_{2} } - \ell_{\hat{\theta}_{1}}\right)  
\longrightarrow 
K\left( \mathbb{P}_{\rho} \left| ~ \mathbb{P}_{\rho^{*}_{1}}\right.\right)
\qquad as~ n\to\infty.
$$
For our particular example we used one of the rank one MLE's to compute an approximate value  
$K\left( \mathbb{P}_{\tau} \left| ~\mathbb{P}_{\rho} \right.\right)\approx 11.33$ which gives an estimate 
\begin{eqnarray*}
BIC(1) - BIC(2) &=& -2( \ell_{\hat{\theta}_{1}} - \ell_{\hat{\theta}_{2}})  + \log (n \cdot 3^{4}) ( p(4,1) - p(4,2) )
\\
& \approx & 2\cdot 11.33\cdot 100+ \log(100\cdot 3^{4})( p(4,1)- p(4,2) )\\
&\approx & 2266 -261 = 2005,
\end{eqnarray*}
in agreement with the histogram illustrated in the right panel of Figure \ref{fig.diff.bic}.

In conclusion, for low rank states with eigenvalues which are not very close to zero, the BIC and to lesser extent AIC,  identify the correct rank with high probability, the latter having a tendency to overfit the true model. On the other hand, as we will see in the next section, the BIC may underfit the true model when one or more eigenvalues are small.

\section{Study 2: one ion simulations} 
\label{sec.one.ion}

We have seen that the performance of the model selection criteria depends on the spectrum of eigenvalues of the true state, and on the number of measurement repetitions. To investigate this dependence we performed a statistical experiment with three one-ion states (k=1) of different degrees of purity: a pure state, one with eigenvalues $(0.95, 0.05)$, and the other with eigenvalues $(0.72,0.28)$. For each state we simulated datasets with varying number of repetitions $n= 10,50,100, 250,500$. Table \ref{table.BIC-AIC-oneion} shows the number of times (out of 1000 samples) BIC and AIC choose the correct rank of the state, for all possible choices of states and measurement repetitions. As expected, in the case of the the pure and the mixed states both criteria require a small number of repetitions (of the order of 50) to give the correct answer.  In the case of the almost pure state, we see a clear dependence with $n$: for small $n$ the difference between the log-likelihoods does not off-set the complexity penalty and both criteria choose rank one; at $n=500$ the balance tips in favour of the rank 2 model, with AIC switching faster than BIC, on average.

\begin{table}[h]
\centering
\begin{tabular}{ c c | c | c | c |c| c |}
\cline{3-7}
& &\multicolumn{5}{|c|}{Measurement Repetitions} \\
\cline{3-7}
&  & 10 & 50 & 100& 250 & 500 \\
\cline{1-7}
\hline
\multicolumn{1}{|c|} { \multirow{2}{*}{State 1}} &\multicolumn{1}{|c|}{BIC} & 987 & 990 &994 &  992 & 996\\
\cline{2-7}
\multicolumn{1}{|c|}{} & \multicolumn{1}{|c|}{AIC}   & 945 & 944 &919 &927 & 930 \\
\cline{1-7}
\hline
\cline{1-7}
\multicolumn{1}{|c|} { \multirow{2}{*}{State 2}} &\multicolumn{1}{|c|}{BIC} & 25 & 83 & 183 &  394 & 706\\
\cline{2-7}
\multicolumn{1}{|c|}{} & \multicolumn{1}{|c|}{AIC}   & 77 & 312 & 502 & 802 & 942 \\
\cline{1-7}
\hline
\cline{1-7}
\multicolumn{1}{|c|} { \multirow{2}{*}{State 3}} &\multicolumn{1}{|c|}{BIC} & 384 & 973 & 998 &  997 & 988\\
\cline{2-7}
\multicolumn{1}{|c|}{} & \multicolumn{1}{|c|}{AIC}   & 594 & 992 &998 &997 & 988 \\
\cline{1-7}
\hline
\end{tabular}
\caption{Performance of BIC and AIC model selection for 3 states: pure (state 1), almost pure (state 2), and mixed (state 3). For each choice of number of repetitions, we record the number of times the BIC and AIC select the \emph{correct} rank out of a total of 1000 simulations.}
\label{table.BIC-AIC-oneion}
\end{table}

\begin{figure}[h]
\begin{center}
\includegraphics[height=4.3cm,width=4.3cm]{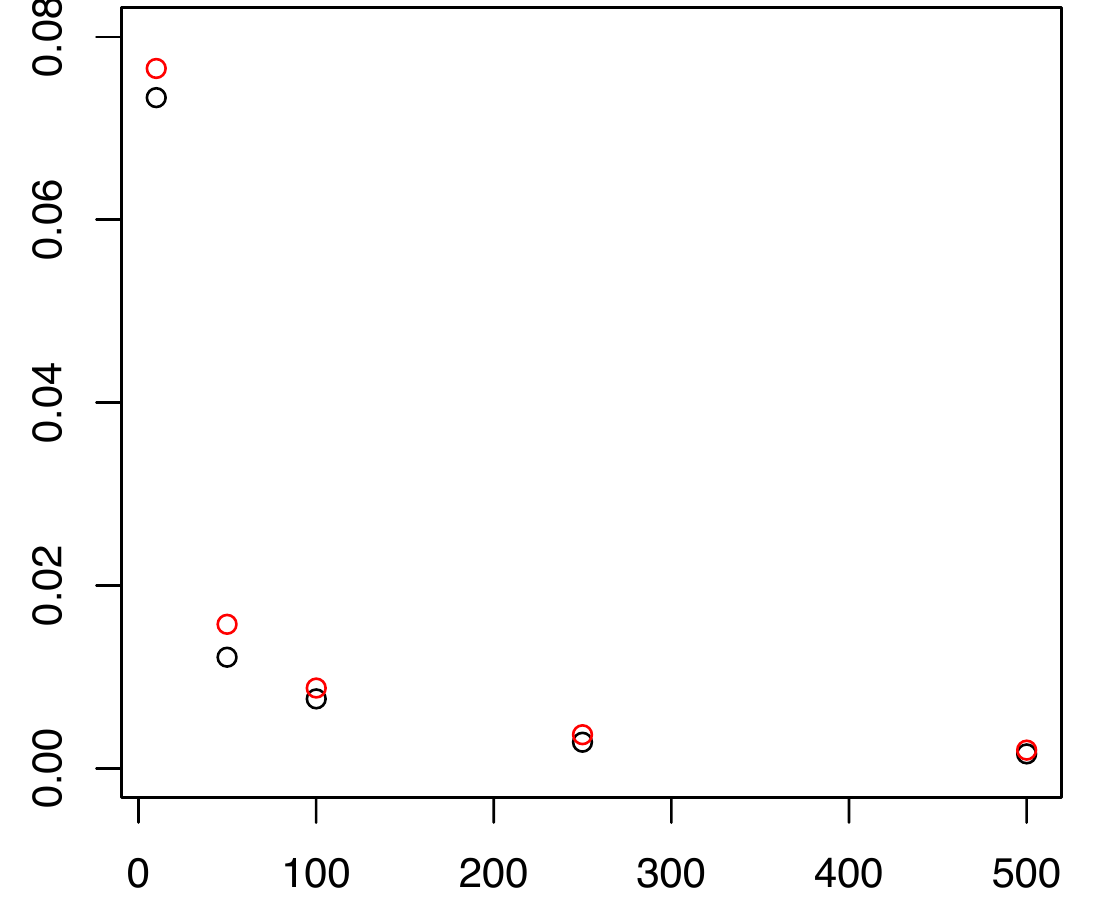}
\includegraphics[height=4.3cm,width=4.3cm]{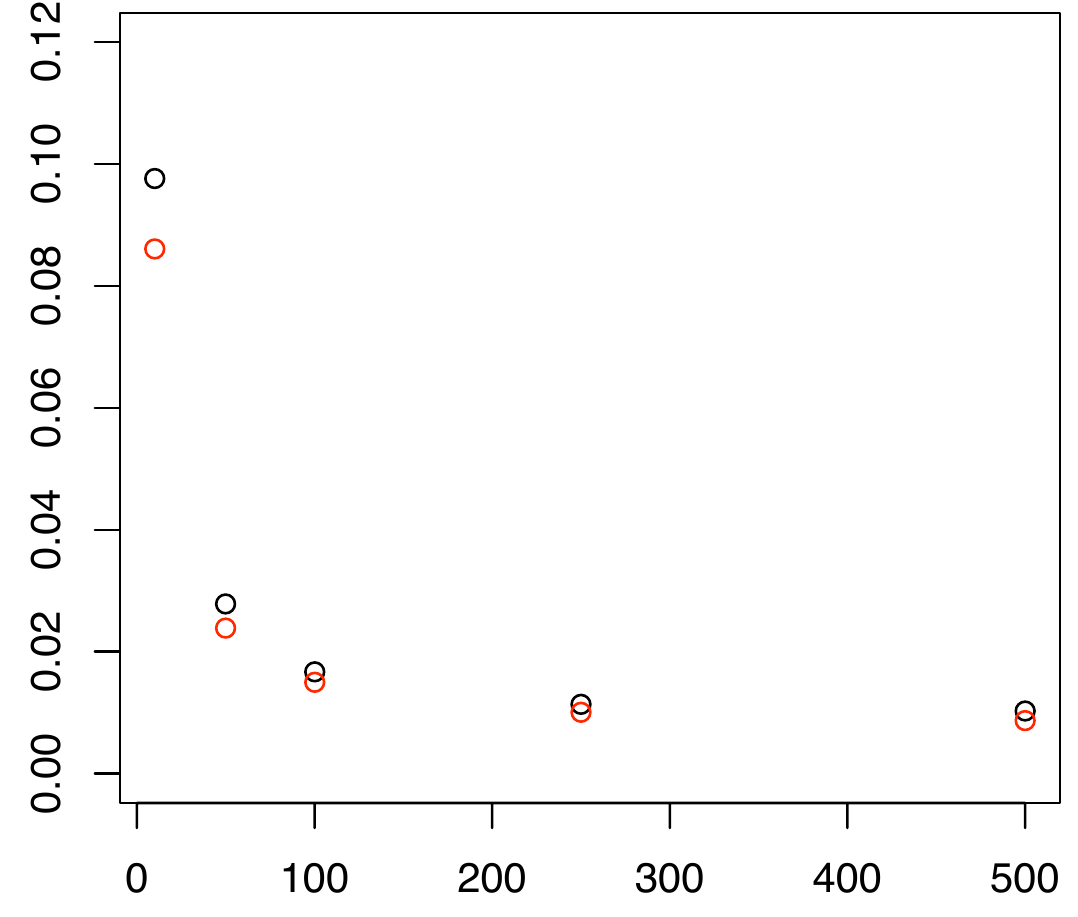}
\includegraphics[height=4.3cm,width=4.3cm]{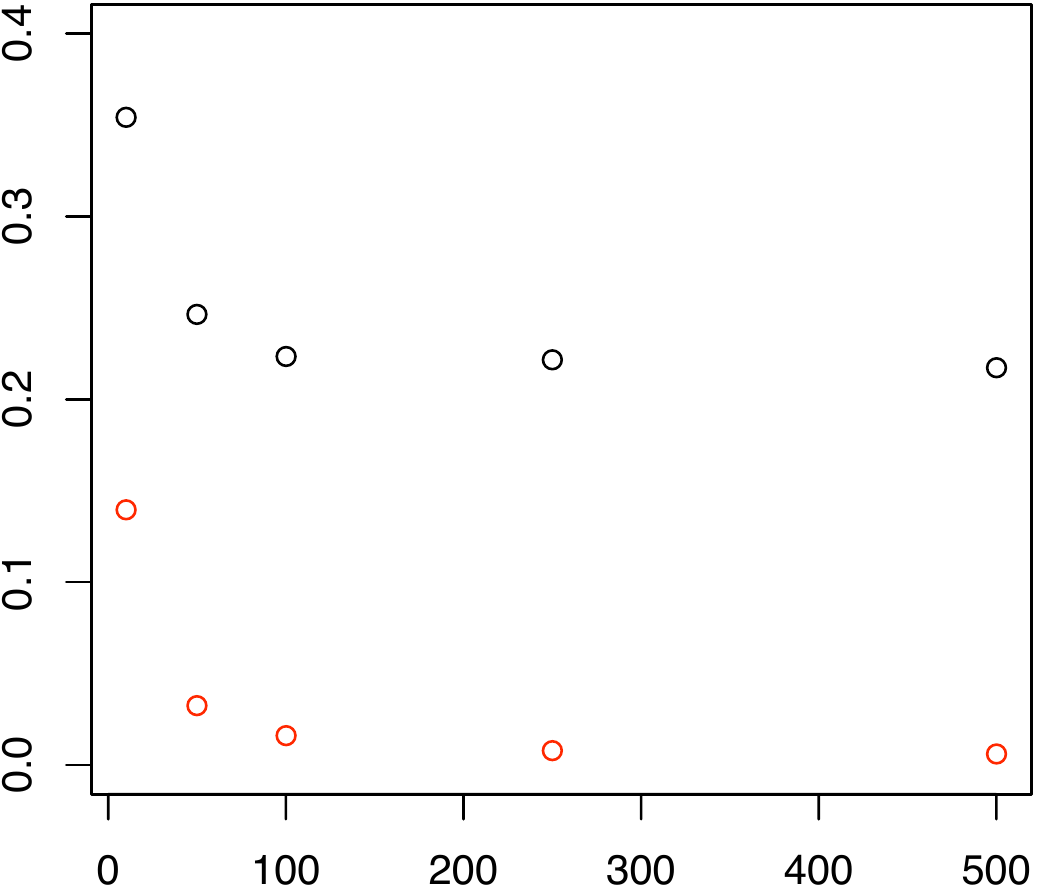}
\caption{Mean square error for rank 1(black circle) and rank 2 (red circle) estimators, as function of the number of measurement repetitions $n=10,50,100, 250,500$. Left: state 1  (pure); Middle: state 2 (almost pure); 
Right: state 3 (mixed).}\label{fig.oneion.mse}
\end{center}
\end{figure}

Figure \ref{fig.oneion.mse} shows the mean square errors (MSE) of the two MLE's
$$
MSE(r) := \mathbb{E} ( \| \hat{\rho}_{r} -\rho\|_{2}^{2} ) , \qquad r=1,2
$$
as a function of $n$  for each of the three states, with the pure state (rank one) estimator in black and the mixed state (rank two) estimator in red. For the pure state (left panel), the rank two estimator has a larger MSE due to the variance contribution from the third parameter, but the relative difference between the two MSE's is small for all $n$. In this case the rank one estimator proposed by both criteria is optimal both from the point of view of parsimony, as well as estimation error. For the mixed state (right panel), the rank one estimator has a large bias which dominates the MSE, while the rank two MSE decreases at rate $1/n$, as expected. At $n=50$ the \emph{relative} difference in risk is significant and  both criteria choose the optimal rank-two estimator. The most interesting case is that of the almost pure state (middle panel). Here we see that the relative difference in MSE is not significant for small and medium number of repetitions ($n=10,50,100$), but for larger $n$ the error of the pure state estimator is dominated by its bias while the variance of the full state estimator becomes very small. This behaviour is picked up by the model selection criteria, which on average switch to the more complex model when $n$ is in the interval between $200$ and $500$.

In conclusion, the study shows that both methods become more sensitive to the true rank of the state as the number of repetitions increases, and the switching point increases (on average) with the purity of the state. As for the estimation error, the switch to the  higher rank model appears to happen in the region where the MSE's of the two estimators starts to diverge from each other, which shows that even if the result is suboptimal for small $n$, the relative difference in errors remains small. Finally, the BIC is more aggressive in selecting the lower complexity model, due to the additional log-factor in the penalty.

\section{Study 3: model selection for 4 ions real data} 
\label{sec.real.data}

\begin{figure}[h]
\begin{center}
\includegraphics[width=6cm]{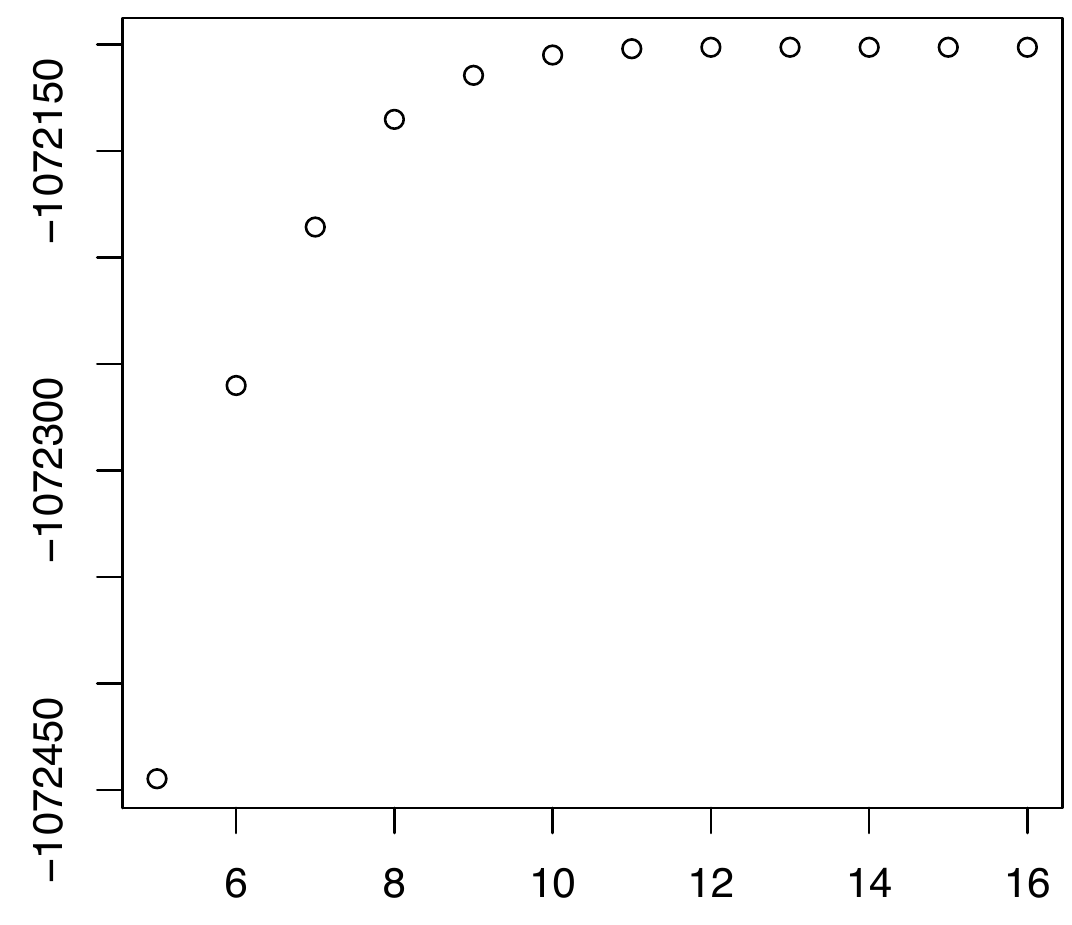}
\caption{Log-likelihood values for the maximum likelihood estimator as a function of rank. }
\label{fig.likelihoods}
\end{center}
\end{figure}

In the third study, we applied the model selection methods to experimental data provided by Rainer Blatt's group from the University of Innsbruck. The aim of the experiment \cite{Barreiro} was to create a particular 4 ions bound entangled state of rank  4 called Smolin state \cite{Smolin}, and the measurement dataset consisted of counts for the $3^{4}$ measurement settings, with a number $n=4800$ of repetitions for each setting. We computed the maximum likelihood estimators $\hat{\rho}_{r}$ for all ranks $r$ between 1 
and 16, and found that the corresponding log-likelihoods reach a plateau at rank 10 (see Figure \ref{fig.likelihoods}) which indicates that the rank 10 model is already sufficiently rich to describe the measurement data. Reinforcing this conclusion, we found that the value of the maximum likelihood for rank 10 (and the subsequent ones) was slightly \emph{larger} than that of that of the maximum likelihood over all states computed with Hradil's iterative method \cite{Hradil}, probably due to the fact that the latter had not reached the true maximum after 1000 iterations.

The values of the AIC and BIC for all ranks are shown on the left side of Table \ref{table.BIC-AIC-realdata}. The two criteria reach minima at ranks  $r=6$ and respectively $r=9$, as a result of the trade-off  between the increasing penalty and the log-likelihood. As expected, the BIC chooses a smaller rank due to its larger complexity penalty, but both methods capture the top 4 eigenvalues of order $10^{-1}$ and a few of the following ones of order $10^{-2}$ which account for experimental imprecision in creating the state. On the right side of Table \ref{table.BIC-AIC-realdata} we listed for comparison the eigenvalues of the rank 10 and full rank estimators, showing perfect agreement in the first two decimal places. We emphasise that results should be taken as an indication that the experimental data is consistent with models whose rank could be chosen somewhere between 6 and 10, rather than answering the ill posed question ``what is the rank of the state''. To make a more informed decision on the final choice of model, one can additionally use different \emph{model testing} procedures such as the Pearson chi-square test  discussed below. As we will see, the various arguments converge towards the conclusion that  the rank 6 estimator may be too conservative, while the rank 10 model already fits the data very well.  
\begin{table}[h]
\centering
\begin{tabular}{ | c  | c | c ||c|c| }
\hline
RANK& AIC &BIC& EIGENVALUES &  EIGENVALUES\\
&&& MLE RANK 10 & MLE RANK 16 \\
\hline
1 &2397395  & 2397722 &2.337  e-01 &2.332 e-01  \\
2 & 2217096 & 2217738&2.290 e-01  & 2.277 e-01 \\
3 & 2170638 & 2171573& 2.258 e-01  &2.253 e-01  \\
4 & 2146295 & 2147502&1.725 e-01 &1.721 e-01  \\
5 & 2145157  & 2146614 & 4.599 e-02 & 4.487 e-02 \\
6 & 2144830 & {\bf 2146515}&2.656 e-02 &2.445 e-02 \\
7 & 2144719 & 2146611&2.385 e-02 & 2.229 e-02\\
8 & 2144652 & 2146728&1.948 e-02&1.884 e-02  \\
9 & {\bf 2144641} & 2146880&1.226 e-02 & 1.155 e-02 \\
10 & 2144648 & 2147028& 1.067 e-02 &1.001 e-02\\
11 & 2144664  & 2147164&0&6.057 e-03 \\
12 & 2144680 & 2147279 &0&2.751 e-03 \\
13 & 2144694 & 2147369 &0&6.779 e-04\\
14 & 2144704  & 2147433&0& 5.278 e-06\\
15 & 2144710 & 2147472&0&2.153 e-06  \\
16& 2144712 & 2147484&0& 1.702 e-16\\
\hline
\end{tabular}
\caption{Left: values of AIC and  BIC for the ML estimators of ranks 1 to 16. The minimum values of the two criteria 
are attained at ranks $r=9$ and respectively $r=6$. \\Right: eigenvalues of the MLE's of rank 10 and 16 in decreasing order}
\label{table.BIC-AIC-realdata}
\end{table}

\subsection{Pearson $\chi^{2}$-test}

As an additional tool for probing the conclusions of the model selection procedures, we recast the problem as that of   testing between the hypotheses
$$
\left\{
\begin{array}{ccc}
H_{0} &=& \text{``the dataset is generated by a state of rank at most r''}
\\
H_{1} &=&\text{``the dataset is generated by a state of rank higher than r''}
\end{array}
\right.
$$
A standard approach to such a problem is  based on using the Pearson $\chi^{2}$-statistic. Following the general procedure described in  appendix \ref{sec.stats}, we consider the rank $ r$ MLE $\hat{\rho}_{r}$ with expected number of counts
$
E({\bf s}|{\bf d}):= n \mathbb{P}_{\hat{\rho}_{r}}({\bf s}|{\bf d}),
$
and define the Pearson $\chi^{2}$-statistic
\begin{equation}\label{eq.pearson}
T(r)= \sum_{{\bf s}, {\bf d}} \frac{ \left(N({\bf s} | {\bf d}) - E({\bf s}|{\bf d})\right)^{2} }{E({\bf s}|{\bf d} )} ,
\end{equation}
where $N({\bf s} | {\bf d})$ are the counts from the real data. Under the hypothesis $H_{0}$,  the Pearson statistic has an asymptotic $\chi^{2}$ distribution with number of degrees of freedom equal to the number of free parameters of the dataset minus the number of parameters of the model
$$
\rm{df}(r):= 3^{4}\cdot (2^{4}-1) - p(r,4).
$$  
Therefore one can define the (asymptotically) level $\alpha$ test 
$$
\left\{
\begin{array}{ccc}
\text{if}~~T\leq  t_{\alpha} &:& \text{accept} ~~H_{0}
\\
\text{if}~~T>  t_{\alpha}  &:&\text{accept} ~~H_{1}
\end{array}
\right.
$$
where the threshhold $t_{\alpha}$ is chosen such that $\mathbb{P}(Y>t_{\alpha})= \alpha$ for a  
$\chi^{2}(\rm{df}(r))$-distributed random variable $Y$. In practice the $\chi^{2}$ approximation works well for pure states 
$(r=1)$, and small rank states which have only a few small eigenvalues. However, if the state has a significant number of small eigenvalues, the distribution of $T(r)$ may differ significantly from the asymptotic $\chi^{2}$ distribution. We will not pursue a theoretical analysis here, but instead use \emph{bootstrap} techniques \cite{Young&Smith} to estimate the distribution of $T(r)$ and then perform the test with respect to the bootstrap distribution. The idea of bootstrap is to use the measurement data itself to construct a distribution which (under the hypothesis $H_{0}$) approximates that of $T(r)$, and therefore can be used to define the threshhold $t_{\alpha}$ instead of the $\chi^{2}$ distribution.

\begin{figure}[h]
\begin{center}
\includegraphics[width=6cm]{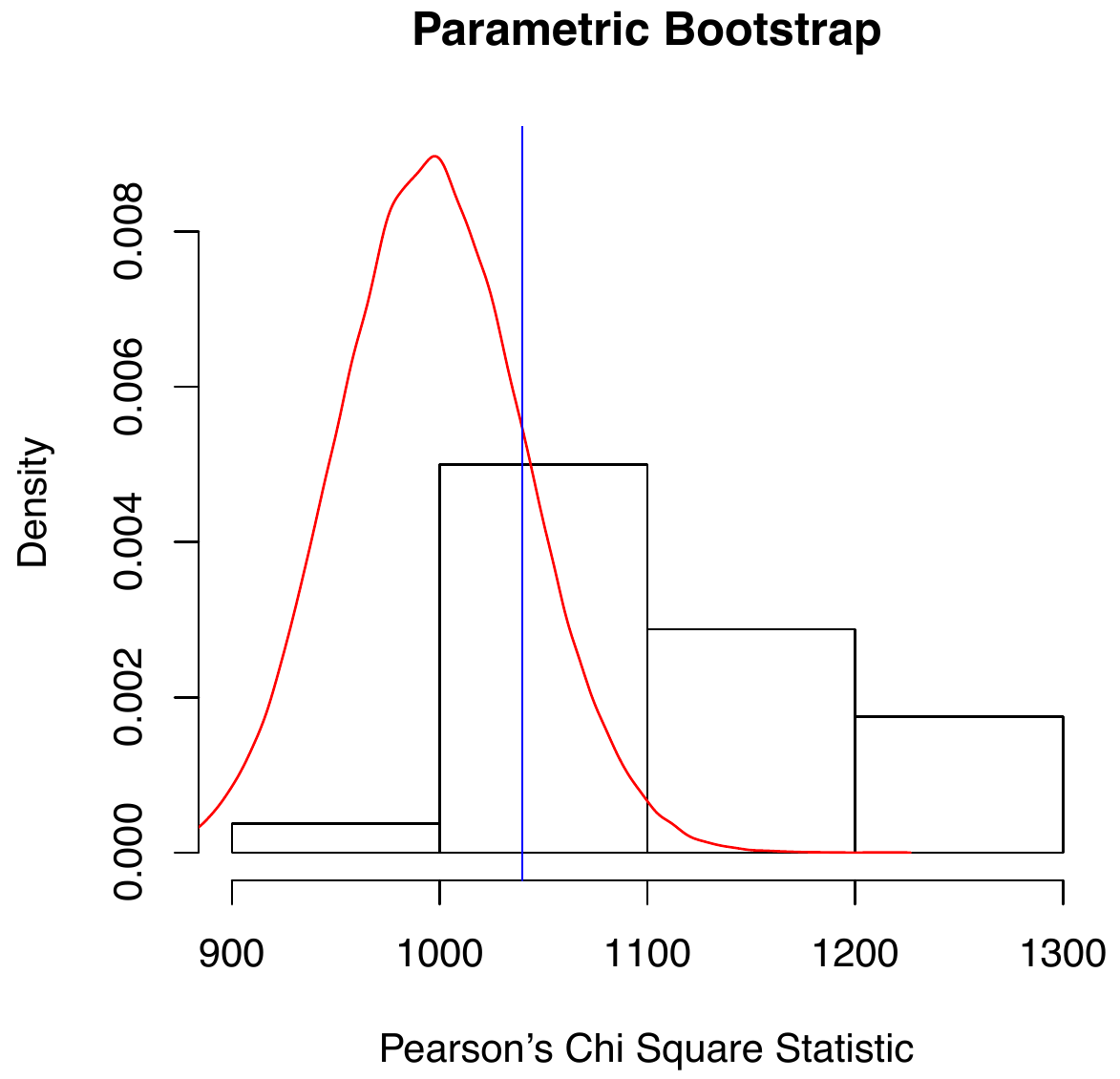}\hspace{2mm}
\caption{Pearson $\chi^{2}$ statistic $T$ (blue line), the limit $\chi^{2}$  distribution (red curve) and the parametric bootstrap distribution for 100 bootstrap samples. The boostrap distribution is shifted with respect to the $\chi^{2}$ due to the fact that the state is close to the boundary of the states space and the asymptotic theory does not hold. Based on the value of 
$T$ we conclude that the hypothesis $H_{0}$ is not rejected for any reasonable significance level. } 
\label{fig.boostrap}
\end{center}
\end{figure}

The bootstrap distributions are constructed as follows:

\begin{itemize}
 \setlength{\itemsep}{5pt}

\item[1)]
Compute the maximum likelihood estimator $\hat{\rho}_{r}$ and its probability distribution $\mathbb{P}_{\hat{\rho}_{r}}({\bf s}|{\bf d})$;

\item[2)]

Generate a large number $N$ of independent datasets from the distribution $\mathbb{P}_{\hat{\rho}_{r}}({\bf s}|{\bf d})$ of the maximum likelihood estimation ;

\item[3)]
Compute the maximum likelihood estimators $\hat{\rho}^{\rm boot}_{1},\dots, \hat{\rho}^{\rm boot}_{N}$ for the bootstrap datasets;

\item[4)]
Compute the Pearson $\chi^{2}$ statistic for each bootstrap sample and its MLE as in \eqref{eq.pearson};

\item[5)]
Apply the $\chi^{2}$ test using the empirical distribution of the boostrap $\chi^{2}$ statistics.  
\end{itemize}  

The results of applying the $\chi^{2}$ test based on the bootstrap distribution to the rank 10 model are illustrated in 
Figure \ref{fig.boostrap}. The value of the test is $T=1039$ is which means that the hypothesis $H_{0}$ is accepted.

\section{Conclusions and outlook}
\label{sec.conclusions}

Statistical inference has become a key tool in interpreting the measurement data in quantum engineering experiments, which require precise, efficient and informative estimation methods. However, standard full state tomography becomes unfeasible for large dimensional quantum systems \cite{Monz}. In this paper we proposed model selection as a general principle for approaching state estimation problems. As in \cite{Gross&Liu&Flammia,Flammia&Gross,Cramer1} the aim is to reduce the dimensionality of the problem by taking advantage of the ``sparsity'' properties of quantum states in realistic experimental settings. The route to this goal is however different. The philosophy in model selection is to try to find the simplest, or most \emph{parsimonious} explanation of the data, by fitting different models (often of increasing complexity)  and choosing the estimator with the best trade-off between likelihood and complexity. Concretely, we looked at the problem of selecting the rank of the estimator, by using two well known methods: Akaike's information criterion (AIC) and the Bayesian 
information criterion (BIC). 
In both cases the fit-complexity trade-off is realised by penalising the log-likelihood of the data 
with a measure of complexity proportional to the number of parameters of the fixed rank model. We have tested AIC and BIC in several real data and simulation studies which we summarise here. 

\emph{Pure states.} We studied the performance of (rank one) ML for pure states and found a very good agreement with the asymptotic predictions based on Fisher information and the efficiency of the ML estimator. More interestingly, we found that the MSE is only slightly larger than the MSE of the \emph{best} possible measurement predicted by quantum version of the asymptotic theory. In particular this rules out the possibility of significantly improving the MSE by means of adaptive measurement design techniques. The (asymptotic) MSE of the full counts dataset was compared to that of the ``coarse grained'' data obtained by estimating the means of the Pauli products corresponding to each measurement setting, as used in compressed sensing algrithms \cite{Gross&Liu&Flammia,Flammia&Gross}. For 4 ions, the latter is an order of magnitude larger than the former due to the loss of information when discarding the full counts statistics.

\emph{Study 1.} For 4 ions states of ranks between 1 and 3 we found that both  AIC and BIC identify the correct rank in 
80\%-90\% of the cases, when the smallest non-zero eigenvalue is not too close to zero. 
The results are explained by using the ML asymptotic theory.

\emph{Study 2.} We analysed the performance of AIC and BIC as a function of the number of measurement repetitions and the purity of the state,  for a toy example consisting of one ion states.  With only a small number of repetitions, both methods identify the correct rank for pure and ``pretty mixed'' states. For an ``almost pure'' state, the model choice switches to rank 2 as the number of repetitions increases. The switching happens roughly at the point where the  MSE of the ``wrong'' rank 1 estimator becomes significantly larger that that of the correct model, indicating that model selection is only slightly suboptimal in terms of the MSE.

\emph{Study 3.} We applied model selection to the 4 ions experimental data provided by Rainer Blatt's group from the University of Innsbruck. The target state of the experiment was an equal mixture of 4 orthogonal pure states, and BIC and AIC selected rank 6 and respectively 9, with both estimators capturing the principle eigenvalues and (some of) the noisy components due to imperfections in the preparation procedure, of the order $10^{-2}$. While the BIC prediction seems too conservative, we find that a rank 10 estimator gives a very good explanation of the data from several perspectives: log-likelihood values, eigenvalues of the estimators, hypothesis testing.

Overall, the numerical results indicate that model selection gives sensible answers, and can be used as an alternative to 
full tomography and compressed sensing. In principle the method works for any state, but is designed to take advantage of the lower complexity of small rank states. The drawback  is the computational 
complexity of finding the MLE over states of fixed rank. Therefore it would be interesting to see whether ideas from the different methods can be combined in a fast, scalable and statistically efficient estimator. A possible future direction is apply model selection to state estimation for other types of models such as classes of matrix product states, and to system identification problems. Another topic of interest is the computation of confidence intervals (error-bars).  Last but not least, there is a need for a deeper theoretical understanding of the quantum tomography statistical model. We mention two important questions: how does the state's proximity to the boundary affect the standard asymptotic theory, and how does the model behave for a large number of ions?  This would hopefully lead to improved estimation algorithms and  information criteria for model selection.

\ack
We thank Rainer Blatt's group for providing us experimental data, and in particular Thomas Monz and Philipp 
Schindler for many fruitful discussions and hospitality during our visits to Innsbruck. MG's research is funded by the EPSRC Fellowship EP/E052290/1. MG and TK acknowledge financial support from the University of Nottingham Additional Sponsorship grant  EP/J501499/1.

\appendix

\section{Pearson $\chi^{2}$-statistic and Wilks' Theorem}
\setcounter{section}{1}
\label{sec.stats}


For reader's convenince we collect here two important results used in the paper. We refer to \cite{vanderVaart,Young&Smith} for more details.

\begin{theorem}[Pearson's $\chi^{2}$ statistic]
Let $X_{1},\dots X_{n}$ be i.i.d. samples from the discrete distribution $\mathbb{P}_{\theta}$ over $\{1,\dots, p\}$, where
$
\mathcal{P}:= \{ \mathbb{P}_{\theta} :\ \theta\in \Theta\subset \mathbb{R}^{m} \}
$ 
is a sufficiently regular model with $\Theta$ an open set. Let $N(i)$ be the number of counts of the outcome $i$ in the sample, and let $E(i)=n \mathbb{P}_{\theta}(i)$ be the expected counts. Then, the Pearson $\chi^{2}$ statistic 
$$
T:= \sum_{i} \frac{(N(i) - E(i))^{2}}{E(i)}
$$
converges in law as $n\to\infty$ to the $\chi^{2}$ distribution with $m$ degrees of freedom.
\end{theorem}

\begin{theorem}[Wilks' Theorem]
Let  
$
\mathcal{P}:= \{ \mathbb{P}_{\theta} :\ \theta\in \Theta= \mathbb{R}^{m} \}
$ 
be a sufficiently regular model and let $\mathcal{P}_{0}$ is the submodel with parameter space 
$
\Theta_{0}:= \{\theta \in \Theta : \theta_{1}=\dots= \theta_{k}=0\} 
$  for some $k\leq m$. Let ${\bf X}:= \{X_{1},\dots, X_{n}\}$ be i.i.d. samples from  $\mathbb{P}_{\theta}$ and let 
$\Lambda$ be the log-likelihood ratio statistic
$$
\Lambda := 
2\left[ \sup_{\theta^{\prime}\in \Theta} \ell_{\theta^{\prime}}({\bf X})  - \sup_{\theta_{0}^{\prime} \in \Theta_{0}} \ell_{\theta_{0}^{\prime}} ({\bf X}) )\right].
$$
If $\theta\in \Theta_{0}$, then $\Lambda$ converges in law as $n\to\infty$ to the $\chi^{2}$ distribution with $k$ degrees of freedom.

\end{theorem}

In both cases, it is essential that the parameter does not lie on the boundary, in order to be able to apply the asymptotic normality theory of the MLE. This condition is violated for states whose rank is strictly smaller than that of the fixed rank model in which they are considered. Therefore care must be taken before applying these results directly, and indeed our results show that the $\chi^{2}$ asymptotics fail in some cases. A more refined asymptotic analysis taking into account the boundary effects will be pursued elsewhere.

\section*{References}

\providecommand{\newblock}{}

\end{document}